\documentclass[sigconf, anonymous=false]{acmart}

\AtBeginDocument{%
  }

\copyrightyear{2024}
\acmYear{2024}
\setcopyright{rightsretained}
\acmConference[CCS '24] {Proceedings of the 2024 ACM SIGSAC Conference on Computer and Communications Security}{October 14--18, 2024}{Salt Lake City, UT, USA.}
\acmBooktitle{Proceedings of the 2024 ACM SIGSAC Conference on Computer and Communications Security (CCS '24), October 14--18, 2024, Salt Lake City, UT, USA}
\acmISBN{979-8-4007-0636-3/24/10}
\acmDOI{10.1145/3658644.3690255}

\usepackage{subfiles}
\usepackage{endnotes}
\usepackage{hyperref}
\usepackage{graphics}
\usepackage{hyphenat}

\usepackage{array}
\usepackage{amsmath}
\usepackage{balance}

\usepackage{caption,tabularx,booktabs}
\usepackage{pifont}% http://ctan.org/pkg/pifont
\newcommand{\cmark}{\ding{51}}%
\newcommand{\xmark}{\ding{55}}%

\usepackage[ruled,vlined,linesnumbered]{algorithm2e}
\SetAlFnt{\small}
\usepackage{algpseudocode}

\usepackage{xcolor,colortbl}
\usepackage{tikz}

\newcommand*\halfcirc[1][1ex]{%
  \begin{tikzpicture}
  \draw[fill] (0,0)-- (90:#1) arc (90:270:#1) -- cycle ;
  \draw (0,0) circle (#1);
  \end{tikzpicture}}
  \newcommand*\fullcirc[1][1ex]{\tikz\fill (0,0) circle (#1);}

\newcommand{\SN}{AutoPatch}
\newcommand{\REC}{Functionality}
\newcommand{\RE}{functionality}
\newcommand{\REL}{functionally}

\usepackage{listings}
\usepackage{makecell}
\usepackage{arydshln}
\usepackage{enumerate}
\usepackage[shortlabels]{enumitem}
\usepackage{diagbox}
\usepackage{rotating}
\usepackage{subcaption}
\usepackage{float}

\definecolor{commentsColor}{rgb}{0, 0.5, 0}
\definecolor{keywordsColor}{rgb}{0.000000, 0.000000, 0.635294}
\definecolor{stringColor}{rgb}{0.558215, 0.000000, 0.135316}
\definecolor{applegreen}{rgb}{0.55, 0.71, 0.0}

\lstdefinelanguage{myLang}
{
	morekeywords={
		bool,
    u32_int
	},
	sensitive=false, % keywords are not case-sensitive
	morecomment=[l]{//}, % l is for line comment
	morecomment=[s]{/*}{*/}, % s is for start and end delimiter
	morestring=[b]" % defines that strings are enclosed in double quotes
}
\lstdefinelanguage{myLangLL}
{
	morekeywords={
		entry,
		alloca,
		\define,
		i32,
		i64,
		call
	},
	sensitive=false, % keywords are not case-sensitive
	morecomment=[l]{//}, % l is for line comment
	morecomment=[s]{/*}{*/}, % s is for start and end delimiter
	morestring=[b]" % defines that strings are enclosed in double quotes
}

\definecolor{greenannoback}{RGB}{230,244,214}

\begin{document}

\title{AutoPatch: Automated Generation of Hotpatches for Real-Time Embedded Devices}


\author{Mohsen Salehi}
\affiliation{%
  \institution{The University of British Columbia}
  \city{Vancouver}
  \country{Canada}}
\email{msalehi@ece.ubc.ca}
\author{Karthik Pattabiraman}
\affiliation{%
  \institution{The University of British Columbia}
  \city{Vancouver}
  \country{Canada}}
\email{karthikp@ece.ubc.ca}

%%\renewcommand{\shortauthors}{Trovato et al.}

\begin{abstract}
  %-------------------------------------------------------------------------------
  Real-time embedded devices like medical or industrial devices are increasingly targeted by cyber-attacks.
  Prompt patching is crucial to mitigate the serious consequences of such attacks on these devices. 
  Hotpatching is an approach to apply a patch to mission-critical embedded devices without rebooting them. 
  However, existing %patching approaches face many challenges, such as requiring restarting mission-critical devices as well as 
  hotpatching approaches require developers to manually write the hotpatch for target systems, which is time-consuming and error-prone.
  
  To address these issues, we propose  
  %addresses these issues by proposing 
  \SN{}\footnote{This paper has been accepted to the ACM CCS 2024 Conference and will appear in the conference proceedings.}, a new hotpatching technique that automatically 
  generates \REL{} equivalent hotpatches via static analysis of the official patches.
  \SN{} introduces a new software triggering approach that supports %a wide range of 
  diverse embedded devices, 
  %and eliminates the need for estimation techniques in hotpatch generation. 
  and preserves the functionality of the official patch.
  In contrast to prior work, \SN{} does not rely on
  hardware support for triggering patches, %writing patches in a new language, and 
  or on executing patches in specialized virtual machines.
  %To this end, we leverage LLVM compiler to instrument applications and generate semantic-preserving hotpatches.
  We implemented \SN{} using the LLVM compiler, and evaluated its efficiency, effectiveness and generality 
  using 62 real CVEs on four embedded devices with different specifications and architectures running popular RTOSes. 
  %The evaluation of \SN{} on popular RTOSes and real-world CVEs demonstrates 
  %its efficiency and effectiveness in addressing security vulnerabilities. 
  We found that \SN{} can fix more than
  % is able to automatically generate hotpatches for over 
  90\% of CVEs, and resolve the vulnerability 
  successfully. 
  The results revealed an average total delay of less than 12.7 $\mu s$ for fixing the vulnerabilities, 
  representing a performance improvement of 50\% over RapidPatch, a state-of-the-art approach.
  Further, our memory overhead, 
  	%s with \SN{} and RapidPatch reveals that, 
  on average, was slightly lower than theirs (23\%).
  Finally, \SN{} was able to generate hotpatches for all four devices without any modifications.
  %\karthik{I wonder if we should tone this down and say comparable?}
  %\karthik{Can we also say something about the real system execution time?}
  %\mohsen{I added this sentence for total overhead, but we removed this total overhead on real applications in Evaluation due to your previous comment:5
  %"I'd expect this to be the case as the overhead is dominated by the instrumentation overhead." Is it OK?}
  %Didn't mention about memory and total incured overhead here:
  %\TC{We also measured the total execution overhead of \SN{} on real applications, 
  %which showed a maximum overhead of 3.9\% in our experiments.}
  \end{abstract}
  


\maketitle

% \documentclass[../main.tex]{subfiles}
% \graphicspath{{\subfix{../images/}}}
% \begin{document}

\section{Introduction}
\label{sec-intro}
%Why we need to fix the vulnerabilities in the embedded devices?
%GPT
%The increasing number of embedded devices connected to the internet, 
%known as the Internet of Things (Embedded), 
%has brought about many conveniences in our daily lives. 
%However, as these devices are used in various industries, from healthcare to power grids, 
%they have also become a target for cyber attacks. 
%Recent examples include the BlackEnergy~\cite{f2014blackenergy} and RIPPLE20~\cite{ripple20} attacks, 
%which have caused physical damage to infrastructure and the exposure of sensitive information (e.g., information of thousands of patients~\cite{medicalvuln}). 
%In 2017, researchers found that certain models of insulin pumps by Medtronic were vulnerable to cyberattacks~\cite{medicalvuln2}, 
%potentially putting patients' lives at risk. 
%This highlights the importance of addressing vulnerabilities in Embedded devices 
%to ensure the safety and security of individuals and critical infrastructure.\par

%Why we need to fix the vulnerabilities in the embedded devices?
%In recent years, %the Internet of Things (Embedded) has revolutionized the way we live and work, 
Embedded devices are used in every aspect of our lives, 
from medical devices and power grids to everyday consumer products.
Due to the use of these devices in different places, especially in critical infrastructures, 
they have attracted the attention of attackers (e.g., RIPPLE20~\cite{ripple20}). 
Successful cyber-attacks on embedded devices can have serious, even life-threatening consequences. Therefore, manufacturers need to prioritize device security by promptly providing patches for vulnerabilities and ensuring regular device updates.

%Therefore, it is crucial for Embedded manufacturers to ensure their devices are secure against cyberattacks, 
%as successful attacks can have serious, even life-threatening consequences.
%Thus, Embedded providers must quickly provide patches for vulnerabilities and ensure devices are updated regularly. 
Studies have found it takes about three months to more than a year for a device to be patched after a vulnerability is discovered~\cite{he2022rapidpatch}. This is because manufacturers face two main challenges for patching vulnerabilities.
First, current updating approaches require rebooting or halting devices 
to fix the vulnerability (\textbf{main challenge 1}), which 
%which due to the possibility of missing the deadline and having irreparable consequences,
 cannot be used in mission-critical places like medical devices.
For example, turning off critical health devices 
(e.g., ventilators, pacemakers) may lead to serious consequences such as the death of the patient.
Second, device manufacturers need to manually write patches for each of their embedded devices, 
%\karthik{Sorry, but is this for each device? We don't automate writing patches, do we?}
which is a time-consuming and error-prone process (\textbf{main challenge 2}). 
\par

To address the first challenge, prior work has proposed hotpatching techniques~\cite{chen2018instaguard,chen2017adaptive}
where the manufacturer can install the desired patch without stopping the device or tasks.
%It should be noted that in this paper, patches generated through hotpatching techniques are referred to as \textit{hotpatches}.
% *Attention: From RapidPatch / Do we need to change the Section 2.5?
% for Android devices and traditional systems, 
However, these approaches require runtime code modification, 
which is not feasible for embedded devices since their code is stored on flash-based ROM.  
%and these approaches need to clear the entire sector and reboot the device. %~\cite{he2022rapidpatch}.  
To alleviate this issue,  
HERA~\cite{niesler2021hera} uses a hardware feature available in the Cortex-M3 and M4 generations of ARM processors 
called FPB~\cite{FPBDoc} (Flash Patch and Breakpoint Unit) to trigger the patch. % and fix device vulnerabilities. 
However, newer generations of ARM (e.g., Cortex-M7)~\cite{NOFPBDoc} and other microcontroller units (e.g., RISC-V) 
do not have this hardware feature, and hence cannot use this approach.
%Prior work has proposed hotpatching techniques~\cite{chen2018instaguard,chen2017adaptive} used in Android devices and traditional systems,
%but little attention has been paid to Embedded devices.
Furthermore, even if the hardware feature is available, 
developers still need to expend manual effort
to compare the patched and unpatched versions of the program and extract the differences 
 to generate the hotpatches.
%\karthik{Sorry, I don't understand. Why do they need to do this?}

%Since most embedded devices are used in sensitive infrastructure, once a vulnerability is discovered, 
%there is an opportunity for attackers to carry out an attack, 
%so providers must quickly fix the vulnerability after it is discovered.
%In recent years, an approach called hotpatching has been proposed, 
%where the manufacturer can install the desired patch without stopping the device.
%In other words, by using the hotpatching method, a program is patched while running without needing to stop or restart (\textbf{main challenge 1}).
%It should be noted that in this paper, patches generated through hotpatching techniques are referred to as \textit{hotpatches}.

%As a result, two hotpatching techniques~\cite{niesler2021hera,he2022rapidpatch} 
%have proposed to fix the security vulnerabilities in real-time Embedded devices. 
%They use FPB~\cite{FPBDoc} (i.e., flash patch and breakpoint unit) hardware to trigger the patch, 
%which is not present in the new generations of ARM (e.g., Cortex-M7)~\cite{NOFPBDoc} and other microcontroller units 
%(e.g., RISC-V). 
% \mohsen{Is it OK that we removed the hardware part since they are hybrid?}
To address the first limitation (i.e., relies on FPB), 
RapidPatch~\cite{he2022rapidpatch} modifies applications to 
%which leverages either hardware or software to trigger the patch.
%If the device does not support the hardware to trigger the patch (e.g., FPB), 
 add a branch instruction (trampoline\footnote{A single instruction, such as a branch or jump instruction, that is added to the application 
%is inserted at arbitrary points in an application to 
to redirect its original control flow.}) at the beginning of each function 
that will execute the desired patch after deployment.
Furthermore, they use the eBPF~\cite{eBPFwebsite} (Extended Berkeley Packet Filter) virtual machine (VM), to execute 
%on the embedded devices, which can execute a 
patches written in a specialized language. 
% *ATTENTION: From RapidPatch
%To fix the vulnerabilities, they support \textit{filter} patches that checks the malicious inputs 
%and \textit{code replace} patches that replace the vulnerable code with a secure patched code. 
%so that device maintainers can use a common patch to fix the same security vulnerability regardless of device architectures. 
However, their approach has two limitations. 
First, using a VM incurs performance and memory overheads, which is often challenging for embedded devices, especially real-time embedded devices.
Second, the real-time operating system (RTOS) developer has to manually rewrite the patch in the eBPF language after writing the patch in C. 
%which is time consuming and they do not want to do it.
%Yet, the second main challenge has not been solved and all the presented approaches require programmer 
This is both time and effort-intensive, especially for larger patches resulting from CVEs. % (\textbf{main challenge 2}). 
%Finally, their approach requires developers to learn the  specialized eBPF language.

%\mohsen{If I want give example, I should say CVE-2020-10062 for example, 
%but the CVE itself is small, but they have to change it and it becomes large, 
%and the explanations of these things have nothing to do with this section.}
%\karthik{Can we somehow quantify this effort or say why it is challenging?}

%In other words, there is no automatic hotpatching method without much overhead for real-time embedded devices 
%that device maintainers can automatically fix device vulnerabilities without having to rewrite the patch 
%in another language.\par
% \mohsen{Why we cannot say "the first automatic hotpatching approach"?}
To address the above challenges, we propose \SN{}\footnote{\SN{} was also presented as a 3-page poster at ACM CCS 2022~\cite{salehi2022poster}.}, an approach for device maintainers to \emph{automatically generate
hotpatches for their embedded devices}, and apply the generated patches {\em without halting or rebooting} these devices.
\SN{} has two main goals: 
%1) No need for special hardware (general hotpatch technique),
1) Eliminate the need for using either a VM or hardware support, 
and 2) Generating hotpatches requiring neither manual effort, nor the use of specialized languages by developers.\par

%We don't say it abstracts the patch.....
%*Attention: We don't say anything about LLVM and LLVM IR. Should we mention here?
\SN{} addresses the above goals through two innovations. 
First, it proposes an instrumentation technique based on static analysis 
%\karthik{This is a strong statement. Shall we rephrase it to most?}
to redirect the control flow and trigger the hotpatch to fix vulnerabilities. 
However, the vulnerable function may have complex code patterns (e.g., dynamic loops), which makes them difficult to analyze statically by \SN{}.
Existing methods~\cite{he2022rapidpatch,xu2020automatic} use either simplification or estimation techniques 
to abstract the effect of the official patch and generate a hotpatch. However, the resulting hotpatch may not %accurately 
reflect the \RE{} of the original patch. 
This is important for safety-critical embedded devices.  
%However, such an approach may not be suitable for embedded devices that are used in sensitive locations.
%One of the primary objectives of \SN{} is to produce 
%a semantically equivalent hotpatch to the official patch without relying on simplification approaches.
%Nevertheless, if the vulnerable function has those complex code (e.g., dynamic or nested loops),
%\SN{}'s static approach for analyzing the function and official patch may not be able 
%to preserve the semantics of the official patch. 
To address this issue, we insert trampolines into \emph{all locations} that are difficult to be analyzed  statically by \SN{}. % (we define this precisely later). 
Thus, \SN{} does not need to analyze the code at those locations to generate hotpatches.

Second, trampolines are in specific locations, while the official patch may be written anywhere in the vulnerable function.
This is because developers need not have any knowledge of where the trampolines are inserted, as per our goals. 
%will fix the vulnerability by modifying vulnerable functions regardless of where trampolines are located. 
%would not consider where the trampolines are inserted when %they develop their patch. 
Therefore, \SN{} should generate the hotpatch so it can be executed by one of the 
% \mohsen{I think the sentence was not correct so I added "by one of the".}
fixed instrumented trampolines, 
and can also have the same effect as the official patch.
To do so, \SN{} automatically analyzes the official patch using a backward static approach,
and generates a hotpatch that is \REL{} equivalent to the original patch, but can be executed 
by one of the instrumented trampolines (goal 2).
%Therefore, \SN{} automatically generates a hotpatch that fixes the vulnerability while being executed 
%by one of the trampolines by using backward static analysis (goal 2). 
%\karthik{Sorry, this makes it seem like the trampoline is executed by the backward static analysis.}
Furthermore, \SN{} operates on the official patch language (i.e., C or C++) to analyze and generate hotpatches, 
eliminating the need for a VM or programming in an alternate language (goal 1).

%To analyze and generate the hotpatch, \SN{} is independent of a specific programming language (goal 1), 
%\karthik{Why is this? We only support C/C++, dom't we. This is a red herring.}
%\mohsen{I thought since AutoPatch works on LLVM IR, we can say every programming language 
%that can be compiled to LLVM IR, AutoPatch can analyze it and generate the hotpatch.}
% and the patch can be written 
%in any languages (e.g., C or C++).

%\karthik{First, explain what filter and replace patches are and why they're important. I removed the text earlier as it was confusing, so you may have to add it back here. }
Similar to previous work HERA and RapidPatch~\cite{he2022rapidpatch, niesler2021hera}, we  
%only 
support two commonly used types of security patches: (1) \textit{filter} patches that check for malicious inputs, and 
(2) \textit{code replace} patches that replace vulnerable code with secure patched code.
These two types of patches are commonly used because they can address a large portion of existing vulnerabilities.
Therefore, in this paper we also consider these types of patches. This excludes any patch that requires changes in the global application's state or data structures from our technique's scope (similar to the other techniques).

\emph{To the best of our knowledge, \SN{} is the first hotpatching framework that %is suitable for real-time embedded devices
%and solves the current problems and 
automatically generates hotpatches without needing either a specialized VM or manual effort.}
%\karthik{More than language, it's programmer support that's important.} 
%\mohsen{I think, "programmer effort" is a better word for this.}
%\par
%
Our contributions are:
%\karthik{There's a lot of repetition and fluff here. The first one is not a contribution, and the third one is more a sales pitch. We have no evidence to back up the "as few trampolines as possible". This needs to be rewritten completely.}
%Is our method similar to RpaidPatch paper contribution? (common patch and don't need to merge source codes)
\begin{itemize}
    \item Propose a new software triggering method for patches by using static analysis techniques  
    for inserting instrumentation into specific locations in the firmware.
    \item Propose an automated static analysis method for analyzing the official patch 
    and generating \REL{} equivalent patches by using backward static analysis.
    \item Design \SN{}, a new hotpatching framework to integrate the software-based triggering with the 
    analysis technique. \SN{} uses the former's output before installing the firmware on the device. %\karthik{We should use the term application rather than firmware.}
    Then, it analyzes the vulnerable function and the official patch to produce 
    the \REL{} equivalent hotpatch. 
    \SN{} is implemented using the LLVM compiler~\cite{llvmcomp}, and hence supports different devices. 
    %for embedded devices.
    \item Evaluate \SN{} on four different RTOSes - 62 real-world CVEs on diverse 
    security vulnerabilities and four different devices in terms of efficiency, effectiveness, and generality. 
    %We compare \SN{}'s performance to the 
    %compared to %state-of-the-art hotpatching technique, 
    %in real-time embedded devices, namely 
    %RapidPatch~\cite{he2022rapidpatch}.
\end{itemize}
   
    The results demonstrate that (1) the hotpatches generated by \SN{} effectively address security vulnerabilities in the CVEs. % in both correctness and efficiency. 
    %I report the entire process here:
    (2) The average runtime overhead incurred by the entire \SN{} execution is negligible (less than 12.7 $\mu s$), resulting in a 50\% average speedup over RapidPatch~\cite{he2022rapidpatch}. 
    This indicates that \SN{} is better than RapidPatch in terms of overhead, despite requiring no manual effort. 
    Further, \SN{} reduces the memory overhead by 23\% compared to RapidPatch.
     %\karthik{Either we should remove this or modify it in the abstract.}
    %\mohsen{Didn't understand.}
    %\karthik{Mention the runtime overhead separately from the analysis overhead}\mohsen{I change "time" to "runtime". I think just mentioning runtime which is the most improatnt overhead in the intro is enough. Otherwise, we should also mention "instrumentation overhead", but they are offline and I think it is not important.} 
    (3) The instrumented code by \SN{} incurs a maximum performance overhead of $3.5\%$ for real applications, when run on the embedded device.
    %\karthik{performance or memory overhead?}
    %\mohsen{Performance. Solved.}
    (4) \SN{} works with different embedded devices  
    without necessitating any changes to its implementation, which demonstrates its generality.

\section{Background}
\label{sec-back}

\subsection{Terminology}
%Throughout the paper, we used terms that will be briefly explained as follows.
%\begin{itemize}
    %\item 
    %1) \textbf{RTOS Developer:}
    %RapidPatch claims they can also work on Bare-metal devices!!! How?!!
    %We also can say that?
    A \textbf{RTOS developer} is an individual or a team responsible 
    for designing and implementing an operating system (e.g., Apache NuttX RTOS~\cite{NuttXcite} and Zephyr OS~\cite{zephyroscite}) 
    that provides real-time capabilities.
 The RTOS developer is also responsible for releasing official patches to fix security vulnerabilities in the RTOS. %\karthik{I used the term official patch. Please check.}
    %\item 
    %2) \textbf{Device Maintainer:}
    
    A \textbf{device maintainer} is an individual or a team that have multiple types of device products.
    These people are responsible for receiving the generated patch (hotpatch) if their device uses a vulnerable 
    version of RTOS, and compiling it using the appropriate compiler and installing it on the target device. %\karthik{Is this hotpatch or official patch?}
    %\item 
    %3) \textbf{Official Patch:}
    %An \textbf{official patch} is a software update released by RTOS developers.
    %These patches are designed to fix bugs and security vulnerabilities (see Section~\ref{cvesec}).

%\end{itemize}

\subsection{Real-Time Embedded Devices}
%One of the types of computers that have been used in various aspects of life in recent years is embedded devices.
%Most embedded devices are implemented for a specific %purpose; 
%therefore, they have limited processing power and memory.
%\mohsen{We don't need to mention about low power and memory in these devices?} \karthik{We don't evaluate power consumption, so no power. As for memory, we're worst off than RapidPatch, so I wouldn't mention it either}
Embedded devices are used in many application areas ranging from industrial control systems to smart homes and medical devices.
%These devices are classified into different classes, 
One class of embedded devices are real-time devices that have a specific deadline for their operation.
%Getting from GPT and HERA
% Based on their requirements, these devices have different strictures on deadlines, 
% which are divided into three categories: hard, firm, and soft.
% Hard real-time systems are systems that must meet a deadline for a task or operation, 
% and failure to do so can have serious consequences, such as the control system for an airplane's engine.
% The systems like traffic control system are firm systems that have deadlines for tasks or operations,
% but the consequences of missing a deadline are not as severe as in hard real-time systems.
% The third category is soft systems (e.g., search engine) that do not have strict timing constraints 
% and missing a deadline may not have any significant consequences.\par 
% \karthik{Which category do we target?}
%HERA and GPT
To manage real-time embedded devices and make sure that tasks do not miss deadlines, 
these devices use special operating systems called real-time operating system (RTOS).
Examples of RTOS include FreeRTOS~\cite{Freertoscite}, NuttX~\cite{NuttXcite}, 
VxWorks~\cite{VxWorkscite}, and Zephyr OS~\cite{zephyroscite,zephyrcite}.
These systems are often used in embedded systems and control applications 
where it is important to have fast and predictable response times.
For example, an RTOS is used in medical devices to ensure 
that patient vital signs are monitored and reported in a timely fashion.

\subsection{CVE and Official Patch}
\label{cvesec}
%Copy from before. 
When a security vulnerability or exposure is discovered, it is publicized and archived 
as a CVE (Common Vulnerabilities and Exposures) in the CVE repository~\cite{mitre}.
%The concept of CVE was introduced by MITRE corporation~\cite{mitre} in 1999 to categorize the vulnerability information 
%that is discovered and make it publicly available.
%In other words, the purpose of CVE is like a free dictionary to organize various vulnerabilities 
%to improve cyber security.
Each CVE is assigned a unique number called CVE numbers or CVE IDs 
so that each vulnerability can be easily referenced and found in the CVE repository.
Listing~\ref{list1} shows an example of a vulnerability discovered in Zephyr OS, the unique number of which is CVE-2020-10021.
\begin{figure}[h]
\begin{lstlisting}[caption=C source code patch for CVE-2020-10021.,label=list1]
1 static bool infoTransfer(void) {
2	u32_t n;
3	n = (cbw.CB[2] << 24) | (cbw.CB[3] << 16) | (cbw.CB[4] <<  8) | (cbw.CB[5] <<  0);
4   LOG_DBG("LBA (block) : 0x%x ", n);
5 +	if ((n * BLOCK_SIZE) >= memory_size) {
6 +		LOG_ERR("LBA out of range");
7 +		csw.Status = CSW_FAILED;
8 +	  sendCSW();
9 +	  return false; }
// ...
}
\end{lstlisting}
\end{figure}
As can be seen, the patch that fixes this vulnerability has been added to the vulnerable function 
in the form of "+" lines (i.e., lines 5 to 9). 
In this paper, patches written by developers are called \textit{official patches}.

%%%Atention: Check them again!! Make sure all things are good and correct.
\subsection{Hotpatching Techniques}
\label{hotpatchback}
Current hotpatching approaches use one of three main methods: 
(1) A/B updating method, (2) instrumenting binary function at runtime, and (3) relocation linked binary libraries.
%\begin{itemize}

\textbf{A/B method:} Patching approaches that use this type of method~\cite{mugarza2020cetratus,Abschemeesp} 
    have two instances of the system at the same time, such as A and B,  
    one of which is running (A) and the other (B) is used for updates. 
    When the B update is completed, the old system A is transferred to the new system B. 
    This method requires storing the information of both instances simultaneously. 
    However, since embedded systems are memory-constrained, these methods are challenging to apply.  
    % embedded devices, these methods can not be used. 
    Furthermore, to transfer from system A to B, there is a need to reboot the device, 
    which is against the purpose of hotpatching.
    For example, the authors use A/B method based on over-the-air (OTA) update for Espressif ESP32 MCU~\cite{Abschemeesp}, 
    %and also in the paper~\cite{Abschemeand} for 
    and Android devices~\cite{Abschemeand}.
    
    Furthermore, many low power embedded devices such as medical devices cannot have multiple devices 
    due to power/form constraints.
    For instance, one cannot install two pacemakers in a single patient’s body. 
    Further, even if a safety critical system has redundant devices executing in lock-step for reliability, 
    rebooting one device will affect the functionality of the entire system (as both devices need to agree on an operation for it to proceed).

\textbf{Instrumenting function:} These methods~\cite{rommel2020global} modify and instrument the instructions of running program.
    Due to the hardware characteristics of embedded devices\footnote{Flash-based ROMs are used in embedded devices to execute code.}, 
    the proposed approaches cannot change the code during the program's execution for adding the trampoline, 
    and it requires clearing the entire flash sector.
    As a result, the device must be halted or rebooted, which is against the purpose of hotpatching approaches.
    Since this type of technique needs to modify the program at runtime to add the patch, they are not applicable for embedded devices. %Further, they also need to halt the device to modify the firmware.
    %\karthik{We already mentioned halting the device.}

\textbf{Relocating linked binaries:} These approaches~\cite{holmbacka2013lightweight} use a feature 
    called dynamic linking. % that exists in non-RTOSes such as Linux and Windows.
    This shared code piece is accessed by the operating system through a data structure that contains symbolic links, which are resolved during execution.
    This feature is used for libraries that want to be shared between multiple applications.  
 %   allowing the system to load these data structures at runtime.
    For hotpatching the system, these approaches add patches by modifying these symbolic links at runtime. 
    However, dynamic linking is typically not supported in real-time embedded devices 
 as it incurs significant performance overhead, and adds uncertainty to the execution time~\cite{niesler2021hera,he2022rapidpatch}. %New for rebuttal

\label{backsec}

\section{Motivation}
\label{sec-motivation}
In this section, we describe the main requirements of hotpatching 
for real-time embedded devices (\ref{sec-challenge}).
Then, we examine current approaches in terms of how they address these requirements (\ref{sec-otherwork}),
and finally how \SN{} addresses them (\ref{sec-solution}). 
%\karthik{Use a macro to refer to the system name if you change it later.}
%\mohsen{Done.}
\subsection{Requirements}
\label{sec-challenge}

%paragraph 1: explain embedded devices and use cases and some examples and say it is very important to fix asap
%Embedded devices are used in various parts of human life, especially sensitive and important places, 
%from transportation to medical devices. 
%They are found in a wide range of industries and applications, including transportation, healthcare, and
%industrial automation. 
%Some important examples of embedded devices include car navigation systems, industrial control systems, 
%and medical devices.
%These devices also have vulnerabilities like other types of systems 
Like any other system, embedded software also has vulnerabilities, 
which can be exploited by attackers to compromise safety (e.g., vulnerabilities in medical devices~\cite{rushanan2014sok}). 
We have identified five requirements that patching approaches for addressing vulnerabilities 
in embedded devices should meet. 
Existing approaches fail to meet these requirements, 
leading to extended periods—ranging from three months to over a year—to address vulnerabilities 
in embedded devices~\cite{he2022rapidpatch}. 
We describe these requirements below.
% \mohsen{I changed this paragraph to address one of the reviewers' comment. 
% Also change the entire section for replacing "challenge" with "requirement". Is it good?}

% %%%%%%%% Commented this part and edited on 17 April for ACM CCS'24:
% %Embedded devices are often used in safety-critical contexts such as medical devices and automotives. 
% Like any other system, embedded software also has vulnerabilities, which can be exploited by attackers to compromise safety (e.g., vulnerabilities in medical devices~\cite{rushanan2014sok}).  % Therefore, it is important to fix vulnerabilities in embedded software in a timely fashion.
% %but due to their importance, they must be fixed immediately after discovering the vulnerability.\par
% %paragraph 2: It cannot because of challenges and rapidpatch shows it takes 6 months or more.
% Unfortunately, it has been found that due to various challenges, it often takes between 3 months to more than one year
% %\karthik{In the intro, you said 6 months to a year. }
% to patch vulnerabilities~\cite{he2022rapidpatch} in embedded devices. %, especially in RTOSes.  
% %(e.g., ZephyrOS~\cite{zephyroscite}, FreeRTOS~\cite{Freertoscite}, and Samsung TizenRT~\cite{tizenrtcite}). 
% We describe the main challenges in updating embedded systems below.

%We describe the main challenges 
%and then discuss about the most important techniques that have these problems. 
%Finally, we state that an approach that has completely solved these main challenges has not yet been proposed.
%All challenges in enumarete.
\begin{enumerate}[start=1,label={\bfseries R\arabic*:}]
    \item \textbf{Memory Constraints}: Embedded devices are designed with cost reduction as a primary consideration, 
    which leads to constraints in different parts such as memory. 
    %Despite these limitation, embedded devices are commonly used in a wide range of applications, 
    %including industrial control systems, consumer electronics, 
    %and IoT devices. 
    Therefore, updating approaches should use the least possible amount of memory  to fix the vulnerability.
    \item \textbf{No Rebooting}: Many real-time embedded devices are used in safety-critical applications like medical devices, 
    and hence there should be no interruption or delay in the execution of main tasks.
    Therefore, the proposed approaches should fix vulnerabilities without interrupting or rebooting 
    the device. 
    \item \textbf{No Special Hardware}: Embedded devices are designed for low power and low cost, and hence the approach should not rely on special-purpose hardware. 
    %they do not have special hardware and usually have the least possible hardware.
    %\mohsen{I didn't delete this part because it doesn't talk about HERA and it's not repetitive. This is an example of the absence of special hardware in devices.}
    For example, there is a hardware feature called FPB~\cite{FPBDoc}
    in the Cortex-M3/M4 series of ARM processors, but 
	later generations of ARM processors do not have it~\cite{NOFPBDoc} (e.g., Cortex-M7) 
    and other microcontroller architectures.
    %Due to their compact size, low cost, and low power consumption, 
    %embedded devices are well-suited for use in portable and battery-powered devices, 
    %as well as in environments where space is limited. 
    Therefore, the proposed approaches should be general and applicable to most devices 
    without the need for special hardware.
    \item \textbf{\REC{} Preservation (Soundness)} 
    The generated patch must be \REL{} equal to the official patch 
    and must be able to have a similar effect on the vulnerable function to fix the security vulnerability. 
    This is because embedded devices are often used in safety-critical contexts, 
    and the generated patch applied to them must be \REL{} equivalent to the original patch. 
    Therefore, patching approaches should not use optimization or estimation approaches that are unsound 
    (e.g., loop summarization algorithms~\cite{xie2017loopster}). 
    %From Vulmet
    \item \textbf{Patch Generation}: 
%    While it is easy for RTOS developers to modify the source code of vulnerable functions to fix bugs, 
 %   it can be difficult to locate the exact locations to make the same changes in binary code. 
    Existing approaches often assume that developers will write patches at %the beginning of 
    functions' entrance~\cite{he2022rapidpatch}. 
	However, in practice, the patch may be written anywhere in vulnerable functions
    (e.g., the middle of the function in CVE-2020-10021, see Section~\ref{cvesec}), 
    and hence developers must manually translate the patch to the beginning of the function (for example). 
%    In order to generate an effective hotpatch, they must have a strong understanding of the official patch 
    %and be able to write a corresponding hotpatch. 
    However, this process is time-consuming and error-prone, and requires significant effort from developers.
    %As a result, it is challenging to provide an automatic hotpatching approach 
    % can find the right location of the hotpatch and modify it in a way that has the same effect as 
    %the official patch and can fix the vulnerability.
\end{enumerate}

\subsection{Other Work}
\label{sec-otherwork}

In this section, we  examine how other approaches address (or not) the requirements outlined in the prior section.
Table~\ref{tab:challenge} summarizes the prior approaches and the requirements they meet. 
%\mohsen{I moved back again this table to the main text. Appendix or main text? (one reviewer suggests move this to Appendix if we don't have space.)}
\emph{As shown in the table, none of the prior approaches meets all requirements entirely.}%We detail below the challenge(s) that each approach fails to address, if any. 

A/B approaches (see Section~\ref{hotpatchback}) store two instances of the system, 
and hence incur large memory overhead, failing to address R1.
Furthermore, these approaches require a reboot to switch between two instances, and hence fail to meet R2. 
As discussed in Section~\ref{hotpatchback}, instrumentation approaches require rebooting of the system, 
and hence fail to meet R2.\par
\begin{table}[ht]
    \centering
    \caption{Summary of other work in terms of the  requirements. \fullcirc: the approach has not met the requirement, and \halfcirc: the approach has partially met the requirement.}
    \label{tab:challenge}
    \resizebox{\columnwidth}{!}{%
    \begin{tabular}{c|c:c:c:c}
    \multicolumn{1}{l|}{\diagbox{Requirements}{Approaches}}    & \begin{tabular}[c]{@{}c@{}}A/B \\Schemes\end{tabular} & \begin{tabular}[c]{@{}c@{}}Instrumenting  \\Function\end{tabular} & \multicolumn{1}{l:}{HERA~\cite{niesler2021hera}} & \multicolumn{1}{l}{RapidPatch~\cite{he2022rapidpatch}}  \\ 
    \hline\hline
    Memory Constraints (R1)                                      & \fullcirc                                     &   -                                                                                                          & -                                                       & -                                                              \\
    No Rebooting (R2)                               & \fullcirc                                         & \fullcirc                                                                                                         & -                                                       & -                                                             \\
    No Special Hardware (R3)                                &  -                                        & -                                                                                                      &  \fullcirc                                                        & -                                                             \\
    Soundness (R4)                                       &  -                                       & -                                                                                                      & -                                                       & \halfcirc                                                           \\
    Patch Generation (R5)                             & \fullcirc                                         &  \fullcirc                                                                                                          &  \fullcirc                                                        &  \fullcirc                                                              \\
    \hline
    \end{tabular}
    }
    \end{table}
%Also, the relocating linked libraries 
%cannot be used for real-time embedded devices due to the lack of dynamic libraries. 
% As discussed in Section~\ref{hotpatchback}, the relocating linked libraries approaches 
% use runtime modification of symbolic links to fix vulnerabilities.
%\karthik{which challenge does this not address? This seems like an implementation issue, btw.}%
%\mohsen{I explained it in the Background Section, but now I exaplain here.} 
% But, this feature is not supported in real-time embedded devices 
% because it has a lot of overhead and also makes the execution time of the program uncertain~\cite{niesler2021hera}. 
%\karthik{Need citation}
% Therefore, the OSes of these devices, 
% such as FreeRTOS~\cite{holmbacka2013lightweight}, do not support this feature. 
% As a result, these approaches are not applicable to these devices.\par

HERA~\cite{niesler2021hera} provides a hotpatching approach for real-time embedded devices. 
% that fixed vulnerabilities 
%despite the limitations of these devices. 
This method leverages FPB to trigger the patch, which is only available in 
Cortex M3/M4 series processors, so this method cannot be used for all embedded devices and hence does not address R3.
RapidPatch~\cite{he2022rapidpatch} uses the branch instruction (trampoline) added at the beginning of each function in the absence of special-purpose hardware.
%But in this method, after the patches are written in C language, 
However, they require the patch to be written in the eBPF language~\cite{eBPFwebsite} so that it can be executed on the device using a VM. This fails to meet R5. 
%\mohsen{We should measure that!}
%\textcolor{red}{The use of VM for embedded devices has a lot of overhead (C1),} also since the patch is written manually, 
%the process is a time-consuming and error-prone process (C5) and needs specialized expertise 
%(familiar with eBPF language).
It should be noted that all existing hotpatching approaches require developers to write the patch manually,  
and as a result do not meet requirement 5 (R5). 
Finally, 
%*Attention: Double check this! ****Attention: I deleted Vulmet.*
RapidPatch inserts trampolines at the beginning of functions, 
%\karthik{This comes out of nowhere. We never discussed Vulmet earlier, did we? Nor is it shown in the table}
%\mohsen{I deleted Vulmet. This work is proposed for Android devices and has this problem.}
%and as the patch location can be anywhere within the function, 
%it utilizes summarization techniques to manage complex instructions (e.g., nested loops), thereby violating C4.\par
%and since the patch may be anywhere in the function, 
and hence uses summarization approaches to deal with complex instructions (e.g., nested loops), which are unsound, violating R4.

\subsection{Our Solution}
\label{sec-solution}
We introduce \SN{}, a hotpatching approach that proposes four solutions (i.e, S1-4) to  
%\karthik{Why should I know about these two phases to understand how \SN{} solves these issues?}
%\mohsen{I deleted this part.}
% to address the above challenges (i.e, C1-5)
%In the following, we discuss how \SN{} 
satisfy each of the requirements (R1-R5).

\begin{enumerate}[start=1,label={\bfseries S\arabic*:}]
    \item \textit{Addressing R1.} Due to memory constraints in embedded devices, \SN{} stores only the generated hotpatches in the device's memory. 
    %*Attentions* Think about that.
    In other words, the memory overhead depends on the official patch, and \SN{} does not need to store entire patched function or any additional information.
    %\karthik{Sorry, but can we say why this is memory efficient? What if the hotpatches are huge?}
    %\mohsen{I say this because some previous approaches require storing the entire patched function. 
    %The largeness of the hotpatch is the result of the largeness of the official patch, 
    %which we cannot do anything about. The previous version had red lines, do you think it is better to say them here again?}
    %unlike the previous approaches that needed to store the entire program in two instances 
    %and transfer the execution process from one instance to another instance, 
    %which induces a high overhead on memory and power consumption (C1).
    \item \textit{Addressing R2 and R3.} 
    We propose a software-only technique to address the issue of rebooting or halting the device 
    to fix vulnerabilities (see Section~\ref{subsec-pre}). 
    This technique involves inserting trampolines in various required locations, 
    allowing for the triggering of generated hotpatches to fix vulnerabilities while the device is still running (R2).
    Furthermore, using this technique, \SN{} does not need special-purpose hardware for triggering the patches (R3). 

    %First, this method is scheduled with the lowest priority, 
    %and as a result, tasks with higher priority can interrupt AutoPatch and be executed. 
    %To avoid rebooting or halting the device to fix the vulnerability, 
    %we propose a new software technique for triggering patches.
    %\mohsen{I wrote the preliminary section, but I also need to discuss about it here and in Design section.} 
    %In this technique, AutoPatch inserts trampolines in different required places 
    %so that they can be used to trigger the generated hotpatches and fix the vulnerability when the device is running.
    %Furthermore, using this technique, AutoPatch does not need hardware for triggering patches, 
    %and with this new software method, we solved second and third challenges.
    %Towards this end, we analyzed 60 real CVEs and determined important locations where patches 
    %are usually inserted (see Section~\ref{sec-pre}). 
    %Based on these locations, AutoPatch inserts trampolines to functions before 
    %the RTOS is installed on the device. 
    %Therefore, in the future, if a vulnerability is found in one of the functions, 
    %AutoPatch will use the most suitable trampoline for triggering the patch. 
    %As a result, AutoPatch does not need hardware for triggering patches, 
    %and with this new software method, we solved second and third challenges. 
    \item \textit{Addressing R4.} 
    Another benefit of our software triggering approach is that \SN{} can generate a \REL{} equivalent hotpatch
    based on the official patch to fix the vulnerabilities.
    Therefore, since \SN{} inserts trampolines in all instrumentation locations (e.g., after nested loops), 
    it does not need to use estimation or summarization approaches~\cite{xie2017loopster} 
    to generate the hotpatches.
    %Another benefit of our new software approach is that 
    %\SN{} no longer needs to use estimation or summarization approaches~\cite{xie2017loopster} 
    %to generate hotpatches. 
    %\karthik{You make this seem like a side-effect of your work rather than a design choice, which is not good. Perhaps you can explain what our main insight here is.}
    %\mohsen{Why is this a side-effect? This is a good point of my approach. I explained in Challenge 4, previous methods use estimation techniques for generating hotpatch that is not good for real-time embedded devices since they use in critical locations. We need accurate patch instead of estimated patch.}
    %Because \SN{} adds trampolines in necessary locations (e.g., after nested loops), 
    %it does not require estimation or optimization methods, and it can generate the accurate hotpatch. 
    %\karthik{What do you mean by accurate?} 
    %\mohsen{I mean, the patch that is generated is semantically exactly like the official patch, 
    %and no estimate was used to make it 
    %(e.g., estimate the number of executions of the loop inside the vulnerable function)}
    %based on the official patch.
    \item  \textit{Addressing R5.}
    We proposed the first automatic hotpatching approach for real-time embedded devices. 
    To achieve this goal, \SN{} analyzes the official patch using static analysis, 
    and automatically generates a hotpatch that can be executed by one of the %\mohsen{Remove "instrumented" (instrumented trampolines)} 
    trampolines.
    %\karthik{I think you need to carefully define the terms official patch and hotpatch earlier as you use them multiple times. Also, is the hotpatch only executed by {\em one} of the trampolines?}
    %\mohsen{I added two new sentences (red lines) to Introduction and Background Sections.}
    %To achieve this goal, when a vulnerability is found, the RTOS developer uses AutoPatch 
    %to generate a hotpatch automatically. 
    %For this purpose, AutoPatch has a phase called the analysis phase, 
    %which analyzes the official patch by using backward static analysis 
    %and automatically generates a patch that is semantically equal to the official patch, 
    %which is called a hotpatch, so that it can be called by the most appropriate instrumented trampoline.
\end{enumerate}

\subsection{Assumptions and Adversary Model}
\label{subsec-assumption-adversary}
%*Attention: Make sure these assumptions and attacker models are correct.
\textbf{Assumptions.}
%HERA and RapidPatch
We make four assumptions as follows.
%\mohsen{Change from itemize to normal text.}
%\begin{itemize}
    %\item 
    First, RTOS developers have already written the official patch for the security vulnerability, 
    %\karthik{what's the difference between a programmer and maintainer ? Do you explain this earlier?}
    %\mohsen{An RTOS developer is the person who wrote the RTOS such as ZephyrOS, while the device maintainer is the person who owns the device that uses that firmware. The previous works have not explained this and have only used it, that's why I did not explain it anywhere. Do I have to explain?}
    and after the hotpatch is generated by \SN{}, 
    the device maintainers have secure access to the patches. 
    We assume that \SN{} has access to the source code of official patches as developers are the ones who use \SN{}. However,  
    because \SN{} operates on LLVM IR, it can also work if RTOS developers release patches in LLVM IR. %\item 
    Second, as in the previous techniques~\cite{niesler2021hera,he2022rapidpatch},
    we assume that hotpatches are transferred in a signed and encrypted form, and so we do not consider malicious modifications of the patch itself.  
    %\karthik{We assume the attacker cannot manipulate the patch after it's transferred either, isn't it?}
    %\mohsen{Yes, that is correct. I added the new sentence to this part.}
    Third, we assume attackers cannot modify the patch after it has been transferred to device maintainers.
    %\item 
    Finally, similar to previous work~\cite{he2022rapidpatch,niesler2021hera}, we assume that the 
    official patches written by the developers are correct (see Section~\ref{sec:discussion}). 
    
    %\TC{%Finally, similar to previous work~\cite{he2022rapidpatch,niesler2021hera},
    %official patches that are written are correct and do not have bugs. 
    %\karthik{Is this similar to other work? If so, say it} 
    %Checking whether the patches have bugs is beyond the scope of the paper. }

    %\karthik{What's the rationale? Does prior work also make this assumption?}
    %\mohsen{Yes, they said: "However, similar to the existing approaches [33,
    %34, 50], RapidPatch itself cannot ensure the patches should
    %work properly, which, in general, is decided by the patch
    %developers."}
%\end{itemize}
%\karthik{I suggest organizing this section as a set of bullet points, and explaining each assumption separately.}
%\mohsen{Done.}

\textbf{Adversary Model.}
%HERA and RapidPatch
We assume attackers perform remote attacks on the device and do not have physical or root access to it. 
However, attackers can exploit memory-safety related bugs (e.g., buffer overflows) 
in the device to overwrite parts of the memory, and launch a
run-time attack.
%\karthik{Do they need root access? We should say if not.}
%\karthik{What's the necessary access here? Root permissions?} and execute runtime attacks. 
%\mohsen{I used the attacker model from previous work (HERA and RapidPatch).}
%\karthik{Again, I think we should provide pointers to prior work in terms of adversary models}. 
%\mohsen{Done.}
As embedded devices' firmware is usually tailored by the device manufacturers 
and lacks support for user applications from external sources, 
it is not possible for attackers to introduce harmful activities, such as executing third-party code, 
through user applications.
Prior work~\cite{he2022rapidpatch,niesler2021hera} has made similar assumptions.

\label{motivsec}

\section{Design}
%I didn't say anything about LLVM! Is it OK? Because RapidPatch say eBPF and ...

In this section, we present the system design of \SN{}, 
the first automatic hotpatching approach for real-time embedded devices
by addressing the issues discussed in Section~\ref{sec-motivation}.
We begin with an overview of \SN{} and its key innovations (Section~\ref{overviewsec}), 
before moving on to describe the different phases of the system in detail (Section~\ref{patchinsec} and Section~\ref{patchasec}).

\subsection{Overview of \SN{}}
\label{overviewsec}

\begin{figure*}[ht]    
	\centering
	\includegraphics[scale=0.7]{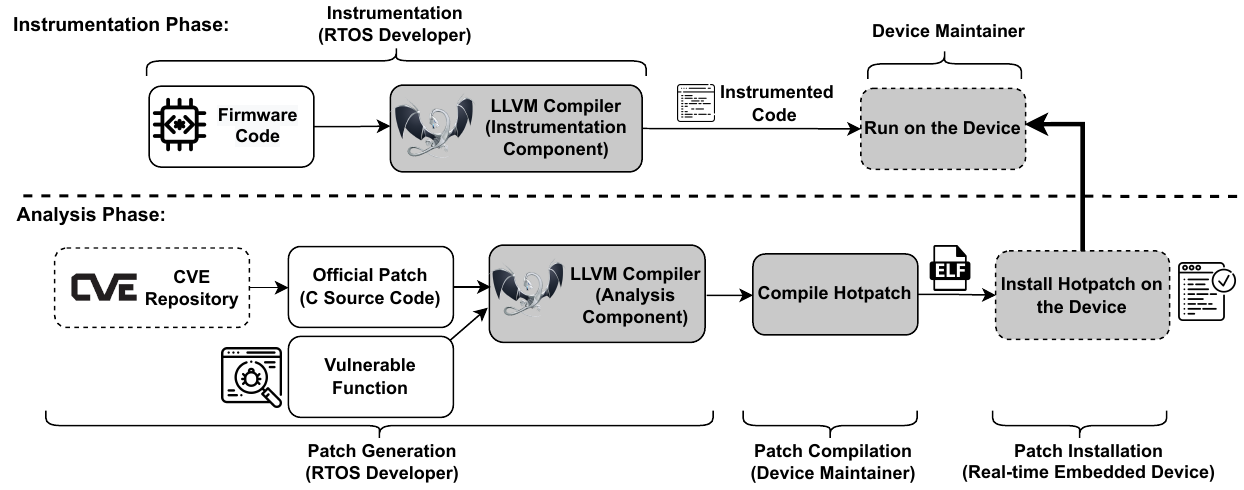} %It was 0.49 (CCS submission, change to )
	\caption{Overview of \SN{}'s operation, including both the instrumentation phase (top) and analysis phase (bottom). }
	\label{OverviewA}
\end{figure*}

Figure~\ref{OverviewA} shows the workflow of \SN{}, 
which consists of two main phases: (1) the instrumentation phase and, (2) the analysis phase.  
Our approach aims to derive and execute \REL{} equivalent hotpatches without relying on hardware or summarization approaches, 
which we have discussed in Section~\ref{sec-motivation} (see S2 and S3). 
To achieve these goals, we propose two innovations. The first innovation is a novel software triggering method that involves \emph{identifying appropriate locations for instrumentation}, and inserting trampolines 
in those locations. We use this method in the instrumentation phase. 
The second key innovation of our approach is the use of \emph{backward static analysis} 
to automatically generate \REL{} equivalent hotpatches based on the official patches to address security vulnerabilities, 
as well as the integration of the hotpatching process with the RTOS to ensure 
that it does not interrupt the execution of other tasks (see S4 in Section~\ref{sec-solution}). 
This is done in the analysis phase, 
which includes an \textit{analysis component (AC)} to generate corresponding patches for the RTOS. 
%To use AC, RTOS developers first write an official patch to fix the vulnerability, 
%and then run AC with two inputs: (1) The official patch, including its location in the vulnerable function, 
%and (2) The vulnerable function. 
%As discussed earlier (Section~\ref{sec-intro}), it is not possible to directly add the official patch to the embedded devices 
%without rebooting it.
%Therefore, AC analyzes the inputs using backward static analysis to generate a hotpatch 
%that can be executed by the most appropriate trampoline at runtime, 
%which is the nearest trampoline to the official patch.
After the hotpatch is generated using \SN{}, it is released for device maintainers, to install on their  devices if they use the vulnerable version. 
% if their devices use the vulnerable version of the RTOS.
%See RapidPatch Design section (especially for runtime)
Then device maintainers compile the hotpatch using a suitable compiler for their platforms, and install it on the device (see S1 in Section~\ref{sec-solution}). \emph{There is no additional effort needed from maintainers to install the patch.}

After the installation, when the vulnerable function is executed, 
the desired patch is triggered, and the vulnerability is fixed without disrupting 
the normal operation of the device or requiring rebooting. %\karthik{Modified this from effort to talk about rebooting - we mentioned effort earlier. }
%These innovations make our approach unique and different from other existing solutions, 
We discuss the process in more detail in the following sections (Sections~\ref{patchinsec} and \ref{patchasec}).
Note that, for the sake of simplicity, the examples are shown using the C language, while \SN{} is based on LLVM IR. 
%Say here about program statements that we called instructions
Further, to explain our examples, we treat each line of code (i.e., program statement) 
in the C programming language as an instruction. In reality, these would be multiple LLVM instructions.

\subsection{Locations of Instrumentation}
\label{subsec-pre}
%**********I am not sure!!!!!!!
%I need to rewrite the motivation part? or Design (Instrumentation Part)?

%**Attention** Be careful% 
%One of the contributions of this paper, 
% \mohsen{Correct it.}
Unlike previous hotpatching approaches~\cite{xu2020automatic,he2022rapidpatch}, 
\SN{} does not use summarization or estimation approaches for generating patches (see Section~\ref{sec-challenge}).
%As a result, we proposed a new software approach for triggering patches to solve the mentioned challenges. 
The goal of \SN{} is to statically create a \REL{} equivalent hotpatch based on the official patch, and to trigger the hotpatch in the vulnerable function.  
%i.e., their impact on the vulnerable function is exactly the same. 
RapidPatch, a prior hotpatching approach~\cite{he2022rapidpatch}, only adds one trampoline at the beginning of each function, as it assumes that developers manually abstract the effect of the official patch at functions entrances. 
Unfortunately, it is not always possible for \SN{} to statically analyze complex code structures and abstract their effects on the patch. Therefore, we insert trampolines inside/after code structures that are too complex to statically analyze. Trampolines will in turn, trigger the hotpatch that will be created at these locations by AC. %\karthik{Added explanation}
%To this end, we add trampolines at appropriate locations in the vulnerable functions to trigger the hotpatch. 
\par

A straw man solution would be adding a trampoline before each instruction to avoid 
the challenges mentioned, but this solution would incur significant overhead, and is hence not suitable for real-time embedded devices.
\emph{Our main observation is that it suffices to insert trampolines in a few strategic locations}. However, the locations for instrumenting trampolines depend on where the patch is inserted. 
In several cases, the trampoline needed to be inserted at the function entrance to redirect execution flow before the vulnerability could be exploited. 
In other cases, the trampoline needed to be inserted at a specific point within the function, 
such as after function calls. 
Overall, to achieve our goal of generating a \REL{} equivalent hotpatch without relying on estimation techniques, 
we identified four \textit{\textbf{instrumentation locations}} where trampolines should be inserted. 
% We chose these locations based on our own observations of real CVEs 
% where \SN{} neither requires the use of estimation techniques nor imposes a large overhead on the device. 
% These locations allow AC to generate a hotpatch functionally equivalent to the official patch 
% even if the official patch is written in a location other than these instrumentation locations. 
Also, we validated these locations with actual CVEs,  and the results confirmed that these instrumentation locations are sufficient 
to execute hotpatches to fix real-world vulnerabilities (Section~\ref{sec-eval-effectiveness}).
%\mohsen{Does this change address to the comments of R1 , 4 , 5?}
These four locations are as follows.
%We describe each of these places.\par

%\mohsen{I changed all "enumerate" to normal text. Is it OK?}
%\begin{itemize}
    %\item 
	\textbf{Function Entrance:}
    The first location for inserting trampolines is at the beginning of every function. 
    This location is crucial for ensuring that any vulnerabilities found 
    at the beginning of the function are patched before the function is executed.
    For example, as shown in Listing~\ref{list1}, a patch was written at the entrance of the vulnerable function to fix a vulnerability in the function.\par
    %\item 
	\textbf{After Function Calls:}
%    Previous hotpatching approaches~\cite{xu2020automatic} to generate a patch when it is after a function call 
%    use approaches such as SVF~\cite{sui2016svf} and summarization methods.
%    While these functions may be complex and the generated patch may not semantically correspond to the official patch. 
%	\karthik{I think we should frame this in terms of why \SN{} needs to insert it after function calls rather than what previous approaches do.}
	%\mohsen{I added this. Is it good?}
%	\TC{The main root of this problem is that in static analysis, 
	Function calls are one of the locations chosen for instrumentation, as 
	it is often difficult to statically determine the effect of the callee function on patch variables, as doing so would require complex inter-procedural analysis.  %\karthik{Rephrased this}
%	and even other functions may be called inside that function.
    Therefore, \SN{} inserts a trampoline after function calls.\par
    %\item 
	\textbf{Inside and After Complex Loops:}
    \textit{Complex loops are loops that either have  conditions (number of executions) that cannot be calculated statically 
    or consist of several nested loops. 
	Otherwise, the loop is considered a \textit{simple loop.}} 
	For instance, when a loop's condition involves a pointer whose value cannot be determined statically, it falls into the category of complex loops.
    %Previous approaches use summarization techniques to estimate the number of loop executions. 
    %In contrast, 
	These loops are challenging for static analysis to accurately ascertain the precise number of loop iterations and the resulting output values.
	Therefore, \SN{} inserts trampolines inside and after these loops to generate a \REL{} preserving hotpatch based on the official patch. 
%	\karthik{Again, explain why NOT in terms of previous approaches.}
%	\mohsen{I added this. Is it OK?}
    %An example of this is CVE-2020-10062, as shown in Listing~\ref{list5p}, 
    %where code snippets were added in two locations of the function to fix the vulnerability (lines with "+").
	\begin{figure}[h]
\begin{lstlisting}[caption=Inserting official patch inside and after a complex loop in CVE-2020-10062.,label=list5p]
1 static int packet_length_decode(struct buf_ctx *buf, u32_t *length) {
2//Allocation Variables
3	do {
4 +  if(bytes >= MQTT_MAX_LENGTH_BYTES){
5 +      return -EINVAL; }
6		// ...
7	} while ((*(buf->cur++) & MQTT_LENGTH_CONTINUATION_BIT) != 0U);
8 + if (*length > MQTT_MAX_PAYLOAD_SIZE) {
9 +    return -EINVAL; }
10	//...
11 }
\end{lstlisting}
		\end{figure}
		
	For instance, CVE-2020-10062, illustrated in Listing~\ref{list5p}, addressed the vulnerability by adding code snippets in two locations of the function (lines with '+'). 
    The code snippets were placed at the beginning of the loop and after the loop respectively. 
    The loop was considered complex as it had a pointer and the number of iterations it executed   
    could not be calculated statically. 
%    In previous approaches, loop summarization was used to generate the patch in such scenarios, 
%    so that the number of patch executions could be estimated. 
%    Since these devices are used in safety-critical contexts, 
%    generating semantic equivalent hotpatch is very necessary, 
%    therefore one of the instrumentation locations to add a trampoline is before and after complex loops.
%\karthik{Prior to this, we've said three times that previous approaches aren't suited for safety-critical systems - we don't need to keep repeating this.}

%\item 
	\textbf{Inside and After Body of Complex Branches:}
    \textit{Similar to complex loops, 
	\SN{} inserts trampolines inside and after the bodies of branches (e.g., if or switch-case statements), 
	when they contain nested branches or their conditions cannot be determined through static analysis (e.g., conditional pointer).}
    In several cases, the official patch is written after a complex 
	% nested 
	'if' statement. In prior hotpatching approaches the entire patched function was executed by a trampoline inserted at the function entrance. 
    But since \SN{} is an automatic approach and its goal is to reduce memory consumption (S1 in Section~\ref{sec-solution}), 
    it inserts a trampoline in these places.

\subsection{Instrumentation Phase}
\label{patchinsec}
The instrumentation phase %produces an instrumented application where the trampolines
inserts trampolines in the instrumentation locations of the firmware functions.
%To achieve our goal (i.e., S2 and S3 in Section~\ref{sec-solution}), 
%We conducted a preliminary study on actual CVEs presented for different RTOSs (e.g., FreeRTOS and ZephyrOS) 
%and identified the necessary locations to insert the trampolines (see Section~\ref{sec-pre}).
To perform this phase automatically, 
\SN{} provides a component called \textit{instrumentation component (IC)} 
that RTOS developers use before releasing their RTOS (e.g., Zephyr OS). 
The task of IC is to statically analyze the functions 
and identify instrumentation locations (e.g., function entrance), where it then inserts a trampoline.
Therefore, IC takes the firmware as an input and outputs the instrumented firmware, 
which contains trampolines in different locations of its functions. 
To do so, IC performs static analysis for each firmware function. 
It examines each instruction within the function, considering the four mentioned instrumentation locations.
If IC identifies a location that matches one of these instrumentation locations, it inserts a trampoline at that point.
%The algorithm outlining the IC process is provided in Algorithm~\ref{alg_ins} in the Appendix.
%Algorithm~\ref{alg_ins} in the Appendix outlines the IC process.

\begin{algorithm}[h]
	\begin{footnotesize}
		\caption{Instrumentation Component Algorithm} 	\label{alg_ins}
		\SetKwInOut{Input}{inputs}
		\SetKwInOut{Output}{output}
		\SetKwProg{instrumentationComponent}{instrumentationComponent}{}{}
		\instrumentationComponent{$(Insts)$}{
			\Input{Function ($Insts = inst_{1} \dots inst_{n}$)}
			\Output{Instrumented Function}
	
			\uIf {Function Entrance}{
				$location := findLocation()$\;
				$insertTrampoline(location)$;
			}
			\Else {
				\ForEach{instruction $inst_i \in Insts$}{%
				
					\uIf{$inst_i.type$ is a $branch\_instruction$}{%
						\eIf{$branch.depth()  > 1$}{
							$beginL := findLocation(branch)$\;
							$endL := findLocation(branch)$\;
							$insertTrampoline(beginL)$\;
							$insertTrampoline(endL)$;
						}
						{ 	
							continue;
						}
					}
					\uElseIf{$inst_i.type$ is a $function\_call$}{
						$location := findLocation()$\;
						$insertTrampoline(location)$;
					}
					\uElseIf{$inst_i.type$ is a $loop$}{
						\eIf{$loop.depth()  > 1$}{
							$beginL := findLocation(loop)$\;
							$endL := findLocation(loop)$\;
							$insertTrampoline(beginL)$\;
							$insertTrampoline(endL)$;
						}
						{ 	
							continue;
						}
					}
				}
			}
			\KwRet{$instrumentedFunction$}\;
		}
	\end{footnotesize}
	\end{algorithm}

Algorithm~\ref{alg_ins} demonstrates how the IC iterates through each instruction (lines 6-25) and checks its type (lines 7, 15 and 18). 
If the instruction is one of the instrumentation locations (e.g., lines 7-12), 
IC uses the $findLocation()$ to obtain the appropriate location for adding the trampoline. 
Once the location is identified, IC calls the $insertTrampoline(Location)$ to add the trampoline to that location.
Otherwise, if the instruction type is not one of the instrumentation locations (e.g., lines 13-14), IC skips it. 
The aforementioned operation is performed for each function in the firmware. % until reaching its end. 
Ultimately, the output of this process consists of the instrumented functions in the firmware code.

% Algorithm~\ref{alg_ins} demonstrates how IC iterates through each instruction (lines 6-25) and checks its type (lines 7, 15 and 18). 
% If the instruction is found to be one of the necessary locations (e.g., lines 7-12), 
% IC uses the $identifyLocation()$ to obtain the appropriate location for adding the trampoline. 
% Once the location is identified, IC calls the $insertTrampoline(Location)$ to add the trampoline to that location.
% Otherwise, if the instruction type is not one of the necessary locations (e.g., lines 13-14), IC skips it. 
% %It should be noted that the trampolines are used to change the control flow of the program and execute the \SN{} manager,
% %whose task is to check the existence of a patch for that function. 

Figure~\ref{instrumentation_example}(a) shows "packet\_length\_decode" function in Zephyr OS. 
The primary purpose of this function is to decode the packet length information that is present in Message Queuing Telemetry Transport (MQTT) packets. 
This information is crucial for the correct processing of MQTT messages.  
%and is required for the reliable functioning of the MQTT protocol. 
%By correctly decoding the packet length, ZephyrOS can ensure that 
%(i.e., the MQTT messages are processed correctly). %\mohsen{This "i.e." is not redundant? I think we can remove it.}
The instrumented function, shown in Figure~\ref{instrumentation_example}(b), is the result of IC execution,
where the green lines represent trampolines. 
\begin{figure}[h]
\begin{minipage}[t]{\columnwidth}
\begin{lstlisting}[title= (a). The C source code function ,label=list2]
1 static int packet_length_decode(struct buf_ctx *buf, u32_t *length) {
2 //Allocation Variables
3	do {
4		if (bytes > MQTT_MAX_LENGTH_BYTES) {
5			return -EINVAL; }
6		// ...
7	} while ((*(buf->cur++) & MQTT_LENGTH_CONTINUATION_BIT) != 0U);
8	//...
9 }
\end{lstlisting}
	\label{list2}
\end{minipage}\hfill

\begin{minipage}{\columnwidth}
\begin{lstlisting}[title= (b). The instrumented function,label=list3]
1 static int packet_length_decode(struct buf_ctx *buf, u32_t *length) {
2 //Allocation Variables
3 %*\colorbox{greenannoback}{Trampoline}*)
4	do {
5   %*\colorbox{greenannoback}{Trampoline}*)    
6		if (bytes > MQTT_MAX_LENGTH_BYTES) {
7			return -EINVAL;}
8		// ...
9	} while ((*(buf->cur++) & MQTT_LENGTH_CONTINUATION_BIT) != 0U);
10  %*\colorbox{greenannoback}{Trampoline}*)
11	//...
12 }
\end{lstlisting}
	\label{list3}
\end{minipage}
\caption{An example of the instrumentation phase for packet\_length\_decode function in Zephyr OS.}
\label{instrumentation_example}
\end{figure}

As seen in Figure~\ref{instrumentation_example}(b), IC has instrumented trampolines for the function 
in three locations: the entrance of the function (line 3), inside the loop (line 5), and after the loop (line 10).
In other words, IC starts to analyze statically from the beginning of the function and 
inserts the first trampoline at the entrance function, after the variable definitions (Line 3). 
Then, this continues the analysis until it encounters a loop.
The loop in lines 4 to 9 has a pointer condition, which IC considers to be a complex loop, as its condition involves pointer artihmetic and cannot be calculated statically. %\karthik{How does it decide this?} \mohsen{I said the reason in Preliminary section, so I refer to it now.}
%\karthik{Added this - please check. }
This results in the insertion of trampolines at the beginning of the loop on Line 5, 
and after the end of the loop on Line 10. 
IC continues this work until it reaches the end of the function.
After the instrumentation process is complete, 
the RTOS developer releases the instrumented firmware.  %instead of the original firmware. 
The device maintainers then install the firmware.

%Note that the trampolines are dormant unless and until a patch is released for this function. 
Note that the inserted trampolines are executed regardless of the existence of the hotpatch, and hence incur a runtime performance overhead (Section~\ref{TED-sec}). The trampolines revert to the normal control flow of the application if no active hotpatch is present.

\subsection{Analysis Phase}
\label{patchasec}
The \SN{} process's next phase is the analysis phase, 
used when a security vulnerability is discovered in the firmware. %\karthik{Why do we give firmware as an example here? We did this earlier as well. Remove it. }
The goal of this phase is to automatically generate hotpatches based on official patches. 
In the instrumentation phase, IC inserted trampolines in instrumentation locations 
%Think about it. I think it is not good!
within functions, while the official patch may have been written anywhere within the function. 
%As a result, the RTOS developer must first determine the best location to trigger the patch, 
%and also understand the semantics of the official patch and modify it 
%so that it can be executed by the chosen trampoline and have the same effect as the official patch. 
%This process can be challenging, time-consuming, and error-prone.
\par

To address this issue, the \SN{} analysis phase is designed to: 
(1) find the best trampoline (Section~\ref{patchgesec})
%\karthik{Perhaps say we explain "best" later}
%\karthik{what do you mean by best?} \mohsen{The mean of "best" is different for each scenario. I explain it for each scenario. What should I do?} 
for triggering the patch, 
and (2) analyze the official patch and automatically generate a 
hotpatch that can be executed by the selected trampoline and have the same effect as the official patch (i.e., \REL{} equivalent hotpatch), 
using backward static analysis. To achieve these goals, the analysis phase consists of three main steps: 
patch generation, patch compilation, and patch installation. 
%In the following, we will provide a detailed explanation of each step in the analysis phase of \SN{}.
The following provides a detailed explanation of each step in the analysis phase of \SN{}.
\subsubsection{\textbf{Patch Generation}}
\label{patchgesec}
As seen in Figure~\ref{OverviewA}, 
after the official patch for the security vulnerability is developed by RTOS developers,
AC receives two inputs from them to generate the hotpatch:
%Changed "vulnerable function" -> "patched function"
1) the patched function, and 
%Changed from "the official patch including the patch and its information (e.g., the patch's location)"
2) the official patch information (e.g., the patch's location within the function).
%\karthik{Why do we need the patch's location separately? Isn't it part of the patch?}
Then, AC statically analyzes these inputs to achieve its goals (i.e., finding the best trampoline 
%\karthik{Again, I suggest not using words like best without defining them.} \mohsen{Discusse about it (above).} 
and generating the hotpatch), 
facing two scenarios depending on the patch location. We explain these scenarios below. \par
%\begin{enumerate}
	%Show example for all scenarios and explain each line what happen.
%\mohsen{I changed all "enumerate" to normal text. Is it OK?}
	%\item 
	\textbf{Outside Body of the Branches (Scenario 1):}
	The scenario pertains to a situation where the official patch is not contained within body of a loop or an `if' statement. 
	An example of such a scenario can be seen in the patch addressing the CVE-2020-10021 vulnerability (Listing~\ref{list1}).
	The \textit{infoTransfer} function after the instrumentation process is shown in Listing~\ref{list2ins}.
	%***ATTENTION:From patch location and backwards ///  OR /// From beginning of function to patch and froward?
	AC starts from the patch location (i.e., line 6 in Listing~\ref{list2ins}) and statically analyzes the function backwards %\karthik{Did you mean backwards?} \mohsen{Corrected.}
	to find the best trampoline.
	The best trampoline is the one closest\footnote{Closest means that the selected trampoline is located before the patch, and has the fewest number of instructions to the official patch compared to other trampolines.}
	%\karthik{We use the term closest in the definition of closest! Closest to what?} 
	%\mohsen{Closest to the patch. Any suggestion? "closest predecessor to the official patch" is Okay?}
	to the patch and its location outside of body of any loops or conditional statements, as determined by the static analysis. %\karthik{Move this earlier} \mohsen{Discusse about it (above).}
%	It should be noted that through the static analysis process, AC can determine whether the patch exists within a loop or if statement, or outside of them. %\karthik{How does it get this knowledge?} \mohsen{Rewrited.}
	%It should be noted that AC has prior knowledge that the patch does not exist within any loops or if statements. 
	Finally, AC selects the trampoline in line 3 as the best trampoline. 
	
%\karthik{You should make sure the figure isn't split across columns, which is what happens now.}
	\begin{figure}[h]
\begin{lstlisting}[caption=Instrumented function with the official patch for CVE-2020-10021 (Listing~\ref{list1}).,label=list2ins]
1 static bool infoTransfer(void) {
2	  u32_t n;
3   %*\colorbox{greenannoback}{Trampoline}*)
4	  n = (cbw.CB[2] << 24) | (cbw.CB[3] << 16) | (cbw.CB[4] <<  8) | (cbw.CB[5] <<  0);
5   LOG_DBG("LBA (block) : 0x%x ", n);
6 +	if ((n * BLOCK_SIZE) >= memory_size) {
7 +		LOG_ERR("LBA out of range");
8 +		csw.Status = CSW_FAILED;
9 +	  sendCSW();
10 +	return false; }
// ...
}
\end{lstlisting}
	\end{figure}

	AC takes the entire patch as input, so it starts again from the patch and goes backwards, analyzing every instruction it encounters. %\karthik{This should be better explained in terms of the CFG.} \mohsen{CFG?}
	If the instruction modifies any of the variables used in the official patch, namely  
	 patch variables (e.g., $n$ and $memory\_size$ in Listing~\ref{list2ins}), 
	AC replaces the instruction values with the corresponding values stored in the patch; 
	otherwise, AC skips the instruction. %As mentioned earlier,
	
	Note that we assume all instructions are in Static Single Assignment (SSA)  form as we work at the level of LLVM IR.
	%\karthik{If you're using instruction, you should explain this is in SSA form and also provide the pseudo-code.}
	This is necessary because \SN{} executes the hotpatch through the selected trampoline, 
	which requires all variables to have the correct values. 
	%\karthik{This is very confusing; what does correct value mean here.} \mohsen{Corrected.}
	In other words, it is possible that between the selected trampoline and the official patch, there  are a number of instructions 
	that change the value of the patch variables. 
	Since \SN{} wants to execute the hotpatch by the inserted trampoline (i.e., changing the location of the hotpatch), 
	AC must check these instructions. If a variable of the patch has been changed by them, it must  
	replace the variable with the changed value, in order to preserve  
	%Therefore, 
	the \RE{} of %generated hotpatch will remain %the same as 
	the official patch.   
%	though it is executed in a different place.
	
	%By modifying the patch in this way, \SN{} ensures that the hotpatch will behave as intended. 
	AC continues this analysis until it reaches the selected trampoline.
	In our example, AC starts from the patch in line 6 ($(n*BLOCK\_SIZE) >= memory\_size$), and works backward.
	Therefore, AC will encounter line 5, and skip it because it has no impact on the patch.
	Then, since the instruction in line 4 changes one of the patch's variables ($n$), 
	the value of $n$ is replaced by the equation on the right-hand side. %***** Attention: I think They don't understand if AC continues it replaces cbw variable for example now.
	As a result, the patch is changed to $((cbw.CB[2] << 24)\textbar{}(cbw.CB[3] << 16) \textbar{} ~ (cbw.CB[4] << 8) \textbar{} (cbw.CB[5] << 0)) *BLOCK\_SIZE >= memory\_size$.
	After AC reaches the selected trampoline in line 3, the hotpatch generation process is finally completed.
	%It should be noted that, for the sake of simplicity, the examples are shown using the C language, while \SN{} is based on LLVM.\par
	%\karthik{This should be mentioned earlier; I suggest though using pseudo-code to describe the patch's operation assuming SSA form.}
	%*********Talk about if there is a "if" or "loop" between patch and trampoline. 
	
	In this scenario, it is possible that there is a loop or conditional statement between 
	the selected trampoline and the official patch. 
	However, since we insert trampolines after complex loops as mentioned in Section~\ref{subsec-pre}, 
	it follows that these loops or conditional statements must be simple if there is no trampoline after them.
	%it is important to note that these loops or conditional statements are simple 
	%\karthik{I'd avoid using words like simple unless you define it. Also, this seems like a circular argument.}
	%\karthik{Also, this seems like a circular argument.}\mohsen{I don't understand!}
	%because if they were complex, there would be a trampoline after them.
	%\karthik{Seems like circular reasoning to me.} \mohsen{Don't undestand.}
	Therefore, when AC encounters a branch (e.g., `if'  statement or loop) while analyzing the instructions, 
	it determines whether the branch modifies any of the relevant values used in the hotpatch. 
	%\karthik{What's a relevant value?} \mohsen{Changed. If we say it like this, won't it be misunderstood? We replace the patch variable if an instruction changes it, and when we continue, the next instruction may replace the changed variable. (Show Example)}
	If none of the patch variables change, it skips the branch.
	%If no relevant value is changed, it skips the branch. 
	Otherwise, AC adds the branch code to the hotpatch. %ATTENTION: What should I say?!

	%\item 
	\textbf{Inside the Body of Branches (Scenario 2):}
	The next scenario is that the official patch is inside the body of a loop or conditional statement. 
	Since the loop or conditional statement can be either complex or simple, 
	%\karthik{I think you need to define complex and simple loops upfront. You keep using these terms without defining them, which is confusing.} \mohsen{I defined them in the Preliminary Section.}
	AC has two modes to handle such scenarios, which are explained in detail below.

	%\begin{enumerate}
		%\item 
		\textbf{Complex Loop/If (Scenario 2-1):} %\karthik{We should formally define complex. What's the difference between simple and complex?}
		%Loop/if has a trampoline
		The IC inserts a trampoline at the beginning and after the complex loop or if statements body in the instrumentation process,
		therefore, if the official patch is inside the complex loop, AC selects the inner trampoline as the best trampoline.
		Then, AC starts from the patch location and moves backwards to reach the selected trampoline.
		%\karthik{Haven't we been doing static analysis all along? What's different now?} \mohsen{Don't understand.} 
		Similarly to the \textbf{Scenario 1}, AC performs the following process:
		it examines each instruction to determine whether it modifies any variable relevant to the patch. 
		If so, it replaces the patch's variable with the value of that instruction. 
		%\karthik{Where does it get the appropriate value from?} \mohsen{Corrected.}
%		This approach ensures that the patch is correctly applied and the hotpatch's variables have correct values.\par
	
		An example of this scenario is shown in Listing~\ref{list5ip}, which is the official patch for CVE-2020-10062.
		The official patch checks the variable $bytes$ at lines 6-7, and $*length$ at lines 11-12.
		As alluded to earlier (Section~\ref{subsec-pre} and \ref{patchinsec}), since the loop has a pointer condition, 
		%\karthik{Where do we say this earlier?}
		\SN{} considers it a complex loop.
		To generate the hotpatch, AC selects the trampoline located in line 5 for the first patch (lines 6-7), 
		and the trampoline found in line 10 for the subsequent patch (lines 11-12).
		Then, since there are no instructions between official patches and trampolines, 
		AC will not change the code of the patches, and hence the hotpatches will be the same as the official patches.
		\begin{figure}[h]
\begin{lstlisting}[caption=Inserting official patch inside and after a complex loop in CVE-2020-10062.,label=list5ip]
1 static int packet_length_decode(struct buf_ctx *buf, u32_t *length) {
2//Allocation Variables
3 %*\colorbox{greenannoback}{Trampoline}*)
4	do {
5  %*\colorbox{greenannoback}{Trampoline}*)
6 +  if(bytes >= MQTT_MAX_LENGTH_BYTES){
7 +      return -EINVAL;}
8		// ...
9	} while ((*(buf->cur++) & MQTT_LENGTH_CONTINUATION_BIT) != 0U);
10 %*\colorbox{greenannoback}{Trampoline}*)
11 + if (*length > MQTT_MAX_PAYLOAD_SIZE) {
12 +    return -EINVAL; }
13	//...
14 }
\end{lstlisting}
		\end{figure}
		
		\textbf{Simple Loop/If (Scenario 2-2):}
		%Loop/if does not have a trampoline
		%I don't find an example for this scenario!
		When the patch is inside a simple `if' statement or loop body, 
		AC considers the entire `if' statement or loop, along with the patch, as a single unit (i.e., patch)
		and applies the same algorithm used in \textbf{Scenario 1}.

Algorithm~\ref{alg_ana} describes the workflow of AC. 
In the first step (Lines 2-8), AC finds the best trampoline based on whether the official patch is in the branch, 
and then stores the instructions between the selected trampoline and the patch in $AInsts$ using the $identifyInsts()$.
Note that AC receives as input the location of the official patch in the vulnerable function from the RTOS developer.
Afterwards (Lines 9-13), AC starts analyzing backwards from the last instruction in $AInsts$, which is the instruction above the patch location ($pLocation$).
If the instruction ($Ainst_i$) modifies a variable of the patch, 
AC replaces that variable with the value on the right side of the instruction; otherwise, AC skips it.
AC continues this analysis until it reaches the selected trampoline (i.e., best trampoline).
Once the trampoline is reached, AC stores the obtained instructions ($PInsts$) in $HInsts$ (Line 14). 
In the future steps (discussed in Section~\ref{patchcsec}), these instructions ($HInsts$) form the hotpatch that will be executed by the selected trampoline.
We summarize the best trampoline chosen for each scenario in  Table~\ref{tab:bestTrampoilne}.

\begin{algorithm}[h]
	\caption{Analysis Component Algorithm}	\label{alg_ana}
\begin{footnotesize}
	\SetKwInOut{Input}{inputs}
	\SetKwInOut{Output}{output}
	\SetKwProg{analysisComponent}{analysisComponent}{}{}
	
	\analysisComponent{$()$}{
		\Input{Vulnerable Function ($Insts = inst_{1} \dots inst_{n}$)\newline Official Patch ($PInsts = pinst_{1} \dots pinst_{n}$)\newline  Patch Location ($pLocation$) \newline
		List of Trampolines ($Tramps = t_{1} \dots t_{n}$)}
		\Output{Hotpatch ($HInsts = hinst_{1} \dots hinst_{n}$)}

		\uIf {$PInsts$ inside a $branch$}{
			\eIf{$branch.depth()  > 1$ || $branch.condition$ has a $pointer$}{
				$bestTrampoline := findTrampoline(Tramps)$; %\Comment{The closest to the patch and inside the branch}\newline
				$AInsts := identifyInsts(bestTrampoline,Insts,pLocation)$;
			}
			{ 	
				$bestTrampoline := findTrampoline(Tramps)$; %\Comment{The closest to the patch and outside the branch}\newline
				$AInsts := identifyInsts(bestTrampoline,Insts,pLocation)$;
			}
		}
		\Else {
			$bestTrampoline := findTrampoline(Tramps)$; \newline
			$AInsts := identifyInsts(bestTrampoline,Insts,pLocation)$;
			
		}
		\ForEach{instruction $Ainst_i \in AInsts = Ainst_{n} \dots Ainst_{1} $}{ 
			
			\uIf{$Ainst_i$ changes a variable of $PInsts$}{
				$PInsts.variable$ <- $Ainst_i$;\newline
				\Comment{Replace the variable with value of $Ainst_i$}
			}
			\Else{
				continue;
			}
		}
		$HInsts := PInsts$ \;
	}
	\KwRet{$HInsts = hinst_{1} \dots hinst_{n}$}\;
\end{footnotesize}
\end{algorithm}

%Furthermore, the algorithm of the AC process is shown in Algorithm~\ref{alg_ana} in Appendix.

\begin{table}
	\centering
	\caption{The definition of the best trampoline for each scenarios. C: Closest trampoline to the patch, 
	D: the best trampoline dominates the exit basic block of the function,
	and S: the best trampoline is in the same basic block of the patch.}
	\begin{tabular}{c|c}
	\textbf{Scenario}       & \textbf{Best Trampoline}  \\ 
	\hline
	Outside Body of Branches (1)   & C , D                                   \\
	Inside Body of a Complex Branch (2-1) & C , S                                   \\
	Inside Body of a Simple Branch (2-2) & C , D                                  
	\end{tabular}
	\label{tab:bestTrampoilne}
\end{table}

\subsubsection{\textbf{Patch Compilation}}
\label{patchcsec}
%***ATTENTION: TALK about address and offset or not?!

%Empty until I figure out how we can compile and complete the patch (e.g., finding the address of the variables).
%Like RapidPatch (Patch Generator)
%*Address??? Offset? Should I say something like RapidPatch and Vulmet?!*
After completing the analysis, the instructions obtained from the previous step 
%along with the official patch,  %**I removed this because in design we said we replaced while in the implementation we add!
are stored in a template function that takes a data structure as an argument 
to receive information from trampolines. 
In other words, when the trampoline is triggered and the hotpatch needs to be executed, 
the values of the vulnerable function should be passed to the hotpatch 
through the hotpatch function's argument.
%\karthik{We need to explain this better}
%\mohsen{Added this.}
%Finally, \SN{} stores the function in a file called \textit{hotpatch} and RTOS developers can release it 
This is released as the hotpatch to fix the vulnerability.
The generated hotpatch for CVE-2020-10062 in the LLVM IR format is shown in Listing~\ref{llvmex} in the Appendix.

% It is important to note that \SN{} operates on the LLVM compiler, 
% which means it saves all debug information during IC and AC, 
% including the base addresses of functions and offset of the variables. 
% This debug information is then utilized during the generation of the hotpatch. 
If device maintainers are using the vulnerable version of the firmware on their devices, 
they would download the hotpatch and compile it using a \textit{cross compiler} 
%\karthik{Why Clang and not other compilers? Do we need to ensure the backend is supported?} 
into an executable file compatible with the target board, allowing them to install it on the board.
% \karthik{How many of these are implementation details? Why should we talk about LLVM here?}

\subsubsection{\textbf{Patch Installation and Execution}}
\label{patchisec}

The last step of the hotpatching process is installing the hotpatch on the board by device maintainers.
Similar to previous work~\cite{he2022rapidpatch,niesler2021hera}, 
%\karthik{cite} 
we assume that device maintainers use an encrypted channel  to transmit the 
%\textit{Executable and Linkable Format} (\textit{ELF})
executable file (hotpatch) alongside 
the patch configuration file including address of the best trampoline for this hotpatch to the board. 
Then, the hotpatch is added to the list of active hotpatches using the \textit{lr} (link register) index, 
which represents the address of the trampoline responsible for executing this hotpatch. 
When the trampoline is executed, it enters the dispatcher section and saves the current values of the vulnerable function. 
If a patch is available, it retrieves the patch using the index from the list of active patches and executes it. 
The stored values are passed through the data frame argument of the hotpatch function. % and executed. 
As each trampoline can have multiple active hotpatches, 
device maintainers can disable them if they are not needed through a secure connection (Section~\ref{subsec-assumption-adversary}).
% \mohsen{Addressing one of the reviewers' comment. But maybe they will say what about attacker can also maliciously disable these patches!}

\begin{figure}[h]    
	\centering
	\includegraphics[scale=0.5]{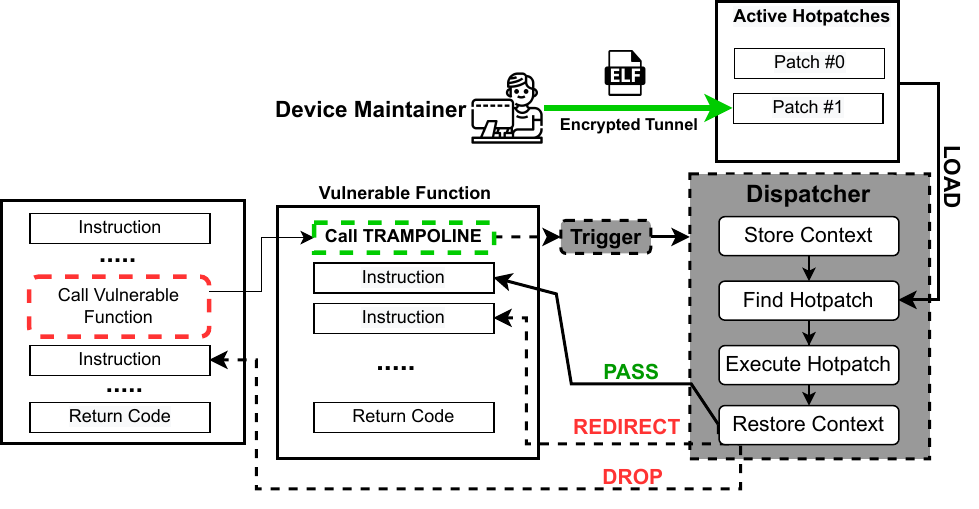}
	\caption{\SN{} Runtime System.}
	\label{RuntimeS}
\end{figure}

Similar to RapidPatch, after executing the hotpatch and based on its result, 
the dispatcher decides to change the program execution flow by changing the return address (\textit{lr}) 
based on one of the three operations: PASS (i.e., resume normal execution of the function), 
REDIRECT (i.e., redirect the program flow to a specific section of the vulnerable function) or 
DROP (i.e., bypass the vulnerable function entirely).
Note that selecting one of these two operations (i.e., REDIRECT and DROP) depends on the official patch.  
%\SN{} does not make the choice. 
%\karthik{Do we need to say this?}
%\mohsen{Reviewer A asked about this. It is not necessary?}
For instance, if a malicious input is detected, the RTOS developer may opt to return to the caller function, 
similar to using DROP in the hotpatch.
Therefore, executing DROP or REDIRECT has no impact on the program's functionality. 
%The overview of this process is shown in the 
Figure~\ref{RuntimeS} shows an overview of the process.

% ****I Should remove from here and said it is similar and explain in two sentences the method (lr and PASS ..) 
%but we don't have VM and we store ELF. 

% After the hotpatch execution, the program flow returns to the normal state (\textit{PASS}) based on the result. 
% Alternatively, it may skip a number of instructions and proceed to a specific instruction (\textit{REDIRECT}), 
% or drop the vulnerable function entirely due to malicious input, 
% \karthik{This is the first time we're mentioning this. What does it mean?}
% returning to the caller function (\textit{DROP}) by changing the lr value.

% Empty until I figure out how we can send the patch and install on the device.
% Also I want to draw a figure (e.g., Figure~\ref{RuntimeS}) for showing how the patch is executed on the device like RapidPatch.

%\mohsen{For figure: It is not completely similar. As we explained, we store ELF and .... It should be mention here?}

% \end{document}
\label{designsec}

\section{Implementation}
\label{sec:implementation}
We implemented \SN{}'s \textit{instrumentation component (IC)} (a total of 1010 LOC in C++ language) 
as an LLVM module pass,
which can be executed on a firmware with different functions~\cite{llvmpass}.
As mentioned in Section~\ref{patchinsec}, the input of this component is firmware code, 
while the output of this component is the instrumented firmware.
Furthermore, the \textit{analysis component (AC)} is implemented as an LLVM function pass with 3100 LOC 
in the C++ language, which is executed on a patched function.
The input of this component is the patched function and patch information (i.e., the patch lines within the function)
and the output is the generated hotpatch, which can be executed by the selected trampoline.
It should be noted that the IC and AC employ static analysis and backward static analysis, respectively, to achieve their respective tasks. 
% The implementation details of \SN{} is described in Section~\ref{sec:implementationA} in Appendix.

% \documentclass[../main.tex]{subfiles}
% \graphicspath{{\subfix{../images/}}}
% \begin{document}

% \section{Implementation Details}
% \label{sec:implementationA}
%?What about the address and offset (Vulmet and RapidPatch)
%?Since we don't have space, I didn't explain what these components do in different situation. 
%?For example, for AC I didn't say what it does when patch is in branch..... Just explain simple scenario. OK?
%\mohsen{I moved Appendix to this location.}
We implemented \SN{} on four types of commonly used connected devices 
%\karthik{What do you mean most used? By whom? Citation?}
%\mohsen{RapidPatch said this? Delete or cite to RapidPatch?}
with different architectures, frequency, and memory storage
to evaluate its effectiveness and generality. They are: 
(1) the nRF52840 development board, (2) STM32\-NUCLEO-F446RE, (3) ESP-WROOM32, and (4) STM32\-NUCLEO-L152RE.
%, as shown in Table~\ref{tab:specification}.
The first two boards are based on the ARM Cortex-M4 but differ in their frequency and memory capacity, 
while the third and fourth boards not only have different frequency and memory 
but also feature distinct architectures (Xtensa and Arm Cortex-M3, respectively).
% compared to the first two.
%\mohsen{Isn't it redundant to say this?}
 Table~\ref{tab:specification} shows the boards' specifications.
%\mohsen{Should we write this sentence in Evaluation Section or this place is good?}
%\karthik{Can we say more about how the boards differ from each other?}
%The architecture of the board is ARMv7-M~\cite{ARMarch} which uses an ARM Cortex-M4 microcontroller with 1 MB flash and 256 KB SRAM. 
%This board was also used by RapidPatch~\cite{he2022rapidpatch}. 
%\karthik{Why do we choose this board? What OSes can it run?}
%\mohsen{No reason, just because one of the boards and kind of the main board for evaluation results that RapidPatch used for evaluation was this board.}
In the following subsections, we explain the implementation details of various components of \SN{} (i.e., instrumentation and analysis components).

\subsection{Instrumentation Phase}
% 1000 LoC cpp 
% llvm module pass containing different functions and 
% develop a script run automatically and get information X,Y .... and the c program and statically analyze the program
% find the necessary locations and insert trampoline
% finally store the new instrumented file in a directory

% We implemented \SN{} \textit{instrumentation component (IC)} (a total of 1010 LOC in C++ language) 
% as a LLVM module pass,
% which can be executed on an application with different functions~\cite{llvmpass}.
As mentioned earlier, the firmware code is given as an input to IC 
and in the first step, 
the firmware is compiled to LLVM IR using the \textit{Clang} compiler~\cite{clangurl}.
Then, IC uses static analysis to iterate through the instructions of the firmware and determine the
instrumentation locations discussed in Section~\ref{subsec-pre} to insert \textit{TRAMPOLINE\_FUNCTION}. 
These locations in LLVM IR are denoted by:
1) \textit{AllocaInst}, 2) \textit{CallInst} and 3) \textit{BranchInst}.
In the case of complex branches (\textit{BranchInst}), encompassing loops or if statements,
IC inserts trampolines both inside and after the branch.
To do so, IC identifies the first instruction in the \textit{header} basic block and the \textit{exit} basic block of that branch,
and inserts trampolines at those locations.
%****ATTENTION: Should I say this below sentence? This is real but it does not have good view in terms of overhead!
It should be noted that if there are multiple exit basic blocks, IC inserts a trampoline at the beginning of each one.
In the code snippet shown in Listing~\ref{listalloc}, IC inserts \textit{TRAMPOLINE\_FUNCTION} at the function entrance
after the allocation instructions (\textit{AllocaInst}).
Finally, IC stores the instrumented file. %as a new file. 

\begin{figure}[h]
\begin{lstlisting}[caption= Inserting Trampoline at the entrance of functions. ,label=listalloc ,language=myLangLL]
define dso_local i32 @infoTransfer() #0 !dbg !50 {
entry:
%n = alloca i32, align 4
%memory_size = alloca i32, align 4
call i64 @TRAMPOLINE_FUNCTION()
}
\end{lstlisting}
\end{figure}

\subsection{Analysis Phase}
% We implemented the process of analyzing the official patch and automatically generating a hotpatch (3100 LOC 
% in the C++ language) 
% as a LLVM function pass, which is executed on a patched function.
The analysis component takes the patched function and patch information %(i.e., the patch lines within the function) 
as input.
Once the function is compiled to LLVM IR using Clang, \SN{} begins static analysis
to find the best trampoline based on the characteristics mentioned in Section~\ref{patchasec}.
Then, AC checks whether the best trampoline to the patch is the previous or next instruction of the patch instructions. 
If so, it indicates that there are no intermediate instructions that could modify the variables involved in the patch. 
In this case, the official patch and the hotpatch are considered identical, and the patch can be executed using the identified trampoline (i.e., best trampoline).\par

% \SN{} begins static analysis from the start of the function 
% and continues until it reaches the patch instructions, searching for the best trampoline based on the characteristics mentioned in Section~\ref{patchasec}.
% Then, AC checks whether the best trampoline to the patch is the previous or next instruction of the patch instructions. 
% If so, it indicates that there are no intermediate instructions that could modify the variables involved in the patch. 
% In this case, the official patch and the hotpatch are considered identical, and the patch can be executed using the identified trampoline (i.e., best trampoline).\par

Otherwise, if the best trampoline is not adjacent to patch instructions, 
\SN{} conducts a static analysis of the official patch. 
It parses the patch and stores all its variables in a data structure. 
Next, it identifies the basic blocks between the patch and the selected trampoline 
and performs a backward static analysis starting from the official patch until reaching the trampoline
to determine instructions affecting patch variables.
Thereafter, AC stores instructions in a new LLVM IR format function, 
whose input is a data frame structure and whose return value is based on one of the three operations 
PASS, REDIRECT or DROP.
Finally, AC saves the function in a \textit{ll} file as a hotpatch,  
which can then be compiled into a file that can be executed on the desired board by using cross compiler.\par

To implement the \SN{} runtime, which includes the trampoline, dispatching, and execution of hotpatches, 
we leveraged the RapidPatch implementation. 
In contrast to RapidPatch (execution part), which stores the eBPF patch code and performs 
a context switch to the eBPF VM for execution, \SN{} stores the hotpatch as an executable file. 
Then, when the dispatcher locates it using \textit{lr}, the corresponding hotpatch function is called with  
the vulnerable function variables passed as arguments.
Therefore, \SN{}  reduces the performance overhead of hotpatch execution over RapidPatch (see Section~\ref{HED-sec} and Section~\ref{subsubsec-comparerapid}).

\label{implementsec}

% \documentclass[../main.tex]{subfiles}
% \graphicspath{{\subfix{../images/}}}
% \begin{document}

\section{Evaluation}
In this section, we describe the evaluation of different aspects of \SN{}.
Specifically, we assessed the effectiveness, efficiency, and generality of the system. 
Effectiveness is the ability of \SN{} to generate patches to fix the vulnerabilities in the firmware, 
while efficiency refers to the runtime and memory overheads of \SN{} (both compile-time and runtime). 
For  generality, we used four different boards with diverse architectures (Table~\ref{tab:specification}) to test if \SN{} 
can work for multiple devices without any modifications.
%\karthik{If this is so, we need to say it explicitly in the earlier section.}
%\mohsen{Added one sentence to Intro section (Contribution).}

\begin{table}[h]
	\centering
	\caption{Devices selected for the experiments.}
	\begin{tabular}{ccccc}
		\hline
		Device             & Arch      & Frequency & Flash & SRAM  \\ \hline
		STM32-L152RE & Cortex-M3 & 32MHz    & 512KB & 80KB \\
		nRF52840           & Cortex-M4 & 64MHz     & 1MB   & 256KB \\
		STM32-F446RE & Cortex-M4 & 180MHz    & 512KB & 128KB \\
		ESP-WROOM32        & Xtensa    & 240MHz    & 448KB & 520KB \\ \hline
	\end{tabular}
	\label{tab:specification}
\end{table}

% We compare \SN{}'s overheads with RapidPatch~\cite{he2022rapidpatch}. 
% Because RapidPatch only reported results for the nRF52840 board, 
% \textcolor{red}{we compared our results of that board with RapidPatch.}
% % we focus on our results for that board in this section. 
% However, we have described the results of \SN{} on the  other boards in Section~\ref{sec:evaluation-extra}.
% %\mohsen{Do you think it's okay to rely too much on RapidPatch and talk about it?}
% %\karthik{Do we also compare with RapidPatch for those boards? If so, we should say that.}

For measuring the compilation times of \SN{} (IC and AC components), 
we used a desktop machine comprising a Intel Xeon CPU E-2224 processor with 4 logical cores at 3.4GHz, 
running Ubuntu 22.04 (64 bit). For the generality of \SN{} %measuring the runtime and memory overheads of \SN{}, 
we used four different boards described in Table~\ref{tab:specification} and different RTOSes.  
%\mohsen{We don't need to mention here we calculate memory and runtime overhead for these boards?}
%it was nrf52840
%\karthik{It's obvious that you'll measure the runtime  overheads on the boards!}

\subsection{Effectiveness}
\label{sec-eval-effectiveness}
We performed this experiment to verify whether \SN{} can successfully fix the security vulnerabilities by generating patches. 
%compared to state-of-the-art hotpatching technique proposed by He et al..
To do so, we utilized RapidPatch's CVE dataset including 62 real-world CVEs, which was collected from the open-access CVE database (NVD~\cite{nvdurl}). 
These CVEs span different RTOSes including Zephyr OS, FreeRTOS and Libraries such as AMNESIA33~\cite{amnesia33} and MbedTLS~\cite{mbedtls}.
More details about these CVEs can be found in their paper~\cite{he2022rapidpatch}.
We analyzed the vulnerable code of these CVEs, 
and confirmed that over 90\% of vulnerabilities (56 out of 62) can be fixed by \SN{}. These results are similar to those obtained by RapidPatch.
The results show that \SN{} effectively mitigates potentially harmful operations that can lead to compromises.

There are two reasons for why \SN{} does not address the remaining six vulnerabilities. 
(1) As discussed in Section~\ref{sec-intro}, 
these techniques cannot generate hotpatches for patches (e.g., CVE-2017-14202 in Zephyr OS) that modify the structures definitions or macros. 
Specifically, out of 62 CVEs, only two CVEs' patches (i.e., CVE-2017-14202 and CVE-2020-13598) modified their macro and structure definitions.
(2) Some patches (e.g., CVE-2020-10064 in Zephyr OS) fixed too many vulnerable functions at the same time.
Attempting to address numerous vulnerabilities through hotpatching all at once might strain the hardware resources~\cite{he2022rapidpatch}.

Further, we manually analyzed the generated hotpatches, and confirmed that the 
\SN{} generated hotpatches mirrored the functionality of the official patches without any bugs.
Also, we asked two developers working in the industry to review the generated hotpatches for functional similarity. 
Both of them confirmed that these hotpatches are functionally similar to the official patches, as far as they could tell.
% \mohsen{I changed the way of saying this sentence.}
Moreover, since the generated hotpatches are not large and complex (averaging around 57 LoC according to Table~\ref{table-CVE-dataset}), reviewing them is not a time-consuming process. 
However, developers can also use different analysis techniques such as static analysis (e.g., LLBMC~\cite{llbmc}) 
% \mohsen{Is it correct? and good to say this?}
to ensure the correctness of the generated hotpatch, but we have not tested that in our evaluation.

\subsection{Efficiency}
\label{sec:eval-efficiency}
%Overhead (Time and memory)
% \karthik{I think we should explicitly say that RapidPatch only used these 12 CVEs}
We evaluated the efficiency of \SN{}'s patching process using 12 real-world CVEs, as listed in Table~\ref{table-CVE-dataset}. Note that RapidPatch also used only these 12 CVEs in their evaluation, and hence we did the same for ease of comparison.

% We evaluated the efficiency of \SN{}'s patching process using the real-world CVEs, as listed in Table~\ref{table-CVE-dataset}. 

\begin{table*}[h]
	\centering
	\caption{\SN{} results on nRF52840 for different CVE types. In the "Triggering" column, we use 
	E: function entrance, F: after function call, 
	I: inside/after if statements, and C: inside/after complex loops. Note that, CVE-2018-16528 includes two official patches.
	The first four columns are from RapidPatch~\cite{he2022rapidpatch}.}
	\resizebox{\textwidth}{!}{%
	\begin{footnotesize}
	\begin{tabular}{c|c:c:c|c:c:c:c:c}
	CVE No.     & OS/Lib        & \begin{tabular}[c]{@{}c@{}}Vulnerability\\Type\end{tabular} & \begin{tabular}[c]{@{}c@{}}Patch \\Lines\end{tabular} & \begin{tabular}[c]{@{}c@{}}Analysis\\Time ($ms$)\end{tabular} & \begin{tabular}[c]{@{}c@{}}Hotpatch Instructions Lines\\ (LLVM IR)\end{tabular} & Triggering                                                         & \begin{tabular}[c]{@{}c@{}}Execution\\Time ($\mu s$)\end{tabular} & \begin{tabular}[c]{@{}c@{}}Memory\\Usage (Bytes)\end{tabular}  \\ 
	\hline \hline
	2020-10021 & Zephyr OS  & Out-of-Bounds Write                                                                                        & 14                                 &    0.8               &  58                                                                   & E        &   0.31                                                                 &   528                                                                                                           \\
	2020-10023 & Zephyr OS  & Logical Bug                                                                                                & 52                                  &     0.6             &     119                                                                 &        C                                                          &    0.20                                                              &      504                                                                                                        \\
	2020-10024 & Zephyr OS  & Instruction Misuse                                                                                        & 11                                    &   0.4             &    34                                                                & E                                                 &   0.26                                                                &  392                                                                                                            \\
	2020-10028 & Zephyr OS  & Lack Sanity Checking                                                                                     & 15                                      &    0.5          &   55                                                                    & E                                                                   &  0.39                                                                   &     400                                                                                                        \\
	2020-10062 & Zephyr OS  & Logical Bug                                                                                               & 38                                      & 0.8             &  32                                                                     &   C                                                                 & 0.48                                                                 &     780                                                                                                            \\
	2020-10063 & Zephyr OS  & Integer Overflow                                                                                           & 12                                      &  0.5            &  28                                                                   & I &  0.36                                                                  & 400                                                                                              \\ 
	\hline
	2018-16524 & FreeRTOS  & Lack Validation                                                                                         & 59                                       & 0.6           &  72                                                                     &   E                                                                 &  0.37                                                                 &  548                                                                                                               \\
	2018-16528 & FreeRTOS  & State Confusion                                                                                           & 6+12                                     &  0.4 + 0.4           &     48 + 46                                                                  &   E                                                                 &        0.36 + 0.41                                                             &  468 + 476                                                                                                          \\
	2018-16603 & FreeRTOS  & Out-of-Bounds Read                                                                                        & 14                                        &    0.5        &  49                                                                     &   E                                                                 &      0.41                                                               &    516                                                                                                        \\ 
	\hline
	2017-2784  & mbedTLS   & Invalid Free                                                                                               & 5                                         & 0.3           &  33                                                                     &  E                                                                  &   0.26                                                                  &    380                                                                                                          \\
	2020-17443 & AMNESIA33 & Lack Packet Checks                                                                                         & 13                                         &  0.4         &     49                                                                  &    E                                                                &  0.31                                                                   &  476                                                                                                               \\
	2020-17445 & AMNESIA33 & Lack Option Checks                                                                                      & 25                                             & 0.9      &  69                                                                     &  C                                                                  &  0.41                                                                  &     560                                                            \\                                            
		\hline
	\textbf{Average}	& & &  & \textbf{0.59} & \textbf{57} &  & \textbf{0.37} & \textbf{535} 
\end{tabular}
	\end{footnotesize}

	}
	\label{table-CVE-dataset}
\end{table*}

\begin{figure}[h]    
	\centering
	\includegraphics[scale=0.7]{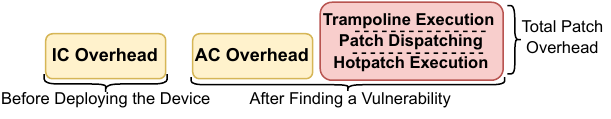}
	\caption{Overview of \SN{} Overheads. Offline and online overheads are shown in yellow and red, respectively.}
	\label{fig-overhead}
\end{figure}

Table~\ref{table-CVE-dataset} shows the results of the evaluation. 
The first four columns of the table (i.e., \textit{CVE No.}, \textit{OS}, \textit{Vulnerability Type} and \textit{Patch Lines}) 
represent the CVE specifications.
We discuss the overhead of each part of the patching process separately,
since they are independent from each other. 
% and we have measured and reported them separately.
%Finally, we discuss the total time that \SN{} takes to fix real-world vulnerabilities on the device 
 Figure~\ref{fig-overhead} shows the \SN{} overheads.
 
There are three components of the overhead: (1) IC overhead, (2) AC overhead, and (3) Total Patch Overhead. 
The first two are incurred offline during compilation, 
while the last one is incurred at runtime during firmware execution. 
The total patch overhead consists of three parts 
% \mohsen{Removed "in turn". It doesn't necessary.} 
(Section~\ref{subsubsec-totaldelay}): (1) trampoline execution overhead, (2) patch dispatching overhead, and (3) hotpatch execution overhead. We explain these below.

%\karthik{Show the name of the overhead in the red box - patch execution delay.}
%\mohsen{Don't understand.}

\subsubsection{Instrumentation Component (IC) Overhead}

% \textcolor{red}{This subsection is done and I measured the overhead by running IC on different functions and reported the average time.}
% \mohsen{Is it appropriate to calculate the average time for the instrumentation compiler (IC) after running it on multiple functions, then multiplying it by the total number of functions (1861 based on RapidPatch) to estimate the time it takes to instrument the entire zephyr. However, as we discussed earlier, it is not feasible for us to compile the entire zephyr to LLVM IR and apply IC on it. Could you please advise if this approach is suitable or suggest any alternative methods?}
%Here, we explain the time overhead for instrumenting process. 
%Also, the memory overhead of trampolines that added to the firmware.
%In the "Adaptability", RapidPatch also explained the overhead of triggering in terms of memory and time very well.
%Since we cannot compile Zephyr OS to llvm ir, I report four CVEs as an example for instrumentation overhead.
\SN{}'s instrumentation phase is executed offline before firmware installation, 
% (Figure~\ref{OverviewA}),  
and is measured on the desktop computer. 
We performed IC on the functions of RTOSes, and calculated the average time required to instrument each function. 
% We performed IC on the different functions associated with our CVE datasets, and calculated 
% the average time required to instrument each function. 
% \mohsen{Maybe the reviewer guessed based on this sentence that we could not compile the entire RTOS. Do you think we should change it and not mention that limitation or is it good as it is?}
% \karthik{We don't say we only instrumented these functions, so I don't think this is the problem. Also, this is for the instrumentation component overhead. }
%??? Using "-time-passes" and see the "Wall Time" for calculating the overhead. We run our pass on a function.
%??? ASK Karthik about this! 
Our results showed that an average of 0.15 $ms$ is needed to identify the instrumentation locations 
in the function and insert the trampoline in those locations.  
%\karthik{Did we also perform the compilation on the embedded board? If not, mention the specs of the computer where you compiled it. }
%\mohsen{No, just on the pc. Where should I add? First of Implementation Sec?}
As an example, we applied IC to the "packet\_length\_decode" function (Figure~\ref{instrumentation_example}) 
associated with CVE-2020-10062. Our LLVM pass successfully identified three instrumentation locations within the function and added trampolines 
at those locations, all within a mere 0.1 $ms$.
%**ATTENTION: TALK About it!
% \TC{In terms of memory overhead, after instrumentation in ZephyrOS, 
% approximately 22 KB of additional space is required, resulting in about 11\% overhead.}
%\karthik{Shouldn't this go to the Trampoline Execution Delay section? This is incurred at runtime, not offline!}
%\mohsen{No, since it belongs to this component. I think it is better place.}

\subsubsection{Analysis Component (AC) Overhead}
%Refer to Table 3
% \textcolor{red}{This subsection is for the analysis component (second phase).
% I want to measure the AC overhead for each of the CVEs and report them separately in Table~\ref{table-CVE-dataset}.}
The next step in the hotpatch generation process involves running AC on the official patch 
and creating the hotpatch via static analysis.
This phase, like the previous one, is performed offline,  % on the RTOS developer system 
and does not impose any overhead on the embedded device.
%To provide an indication of the time required for generating a hotpatch, 
We calculated the average time for generating a hotpatch for the 12 CVEs, as reported in Table~\ref{table-CVE-dataset} 
(column \textit{Analysis Time}). 
Our results show that the average time is 0.6 $ms$, and the  maximum time required for generating a hotpatch is 0.9 $ms$, 
which is negligible.
The \textit{Hotpatch Instructions Lines} and \textit{Triggering} columns %in the table 
respectively represent the number of generated hotpatch
instructions in LLVM IR and the best location within the vulnerable function where the corresponding hotpatch is triggered. 
%\karthik{I'm confused; wouldn't there be multiple such  locations? Also, can we say something about the number of lines of LLVM IR.}
%\mohsen{Explain ....}
%Moreover, the \textit{Triggering} column demonstrates that the considered instrumentation locations are adequate for real CVEs. 
%\karthik{I don't really see how it shows that.}
%\mohsen{I wanted to say since each of the CVEs are called by one of the instrumentation locations (function entrance or inside the loop ...), this shows that those locations were sufficient.}
%\karthik{I suggest removing the claim that it is sufficient as I don't understand how this follows.}
\par

%\TC{Talk about "memory overhead" for different CVEs in the Table and say (like RapidPatch) it needs X KB so it can be used
%in embedded devices:.....}

%These titles are from RapidPatch
\subsubsection{Trampoline Overhead}
\label{TED-sec}
%\karthik{I renamed this section, so we can also discuss the memory overhead here.}
%\mohsen{This section is still about execution of Trampoline not memory and ... }
% \textcolor{red}{This subsection is done. I measured the patch triggering overhead, 
% which is independent of CVE because it consists of a set of fixed number of instructions 
% and can be measured separately.}
%Talk about Fixed patch point. We use the implementation of RapidPatch, Is it OK? What we should do?
%What happened for us that we have multiple triggers in each functions.
% \mohsen{I used the implementation of RapidPatch. Do we need to mention here or in implementation section?}
The software triggering method utilized in \SN{} involves a fixed set of instructions 
for triggering the hotpatch, and storing and retrieving function information. 
Therefore, we can evaluate the overheads of this method independent of the specific patch. %\karthik{Rephrased}
%\karthik{Mention that this overhead is incurred on the embedded device even without a patch.}
%\mohsen{Show the comment in 5.3 Instrumentation Phase in Desing}
%\mohsen{From here, I think the previous version was better. Removing Hardware part. Removing printf.}
To measure the standalone loading delay, 
we executed the software triggering method 20 times on the nRF52840 board 
and calculated the average CPU cycles taken. We found that the triggering approach takes 65 cycles, 
which is approximately 1.01 $\mu s$ for each execution on the board. 
%\mohsen{Since we used RapidPatch implementation, should I remove this sentence that mentions their result? But maybe we can keep their results for hardware triggering?}
%In comparison, the RapidPatch~\cite{he2022rapidpatch} approach incurs 66 cycles of overhead (1.03 $\mu s$) using this method. Therefore, the overall patch triggering delay of \SN{} is comparable with RapidPatch.  
%\karthik{Removed the hardware as it is not comparable with our technique}
%\mohsen{I belive when we say we used their implementation, we don't need to compare them to us!}

%while if they~\cite{niesler2021hera,he2022rapidpatch} leverage hardware methods (i.e., FPB) can take up to 6.5 $\mu s$ (400 cycles) for patch loading. 
%***** ATTENTION: FROM RapidPatch *****
%This in not new! Just ask Karthik we should cite RapidPatch or it is obvious and it is not mandatory?
%I removed the citation but I also changed the time for us. But the time is changed for different strings.
%175,000 for "Hello World, I am AutoPatch!"
%**Attention: Run on "printf" and "Blinky" sample

Furthermore, we evaluated the overhead of the instrumented version of different applications provided by Zephyr OS. We observed a latency increase ranging from 0.05\% to 3.7\% across different scenarios (average of 10 runs each).
RapidPatch incurred 0.05\% to 0.6\% latency overhead in the same scenarios.
%\karthik{I'd provide the corresponding number for RapidPatch rather than the printf function, as it is not directly relevant to us}. 
Thus, \SN{} incurred a higher performance overhead than RapidPatch on the real applications due to the additional instrumentation added by it. This is because RapidPatch only inserts a single trampoline in each function, unlike \SN{}, which inserts multiple trampolines. %depending on the function.  
%\mohsen{Add RapidPatch work on that samples overhead here.}
%\mohsen{printf is a good statement to understand how long a simple program can take, as it affected me when I was reading RapidPatch. Also for hardware.}
%\mohsen{Why do we all have to say our bad results and cover their bad results?}

%Furthermore, after instrumenting ZephyrOS \textcolor{red}{and FreeRTOS} in the instrumentation phase, 
%approximately 22KB \textcolor{red}{and 65KB} of additional memory space is required, 
%resulting in an overhead of 11\% \textcolor{red}{and 8\%, respectively}. 
%\mohsen{In Limitation Section, we said we couldn't compile entire RTOS to LLVM IR! Contradiction? Remove that sentence from Limitation Section?}

Furthermore, instrumenting Zephyr OS and FreeRTOS in the instrumentation phase using \SN{}  
incurred memory overheads of 11\% and 8\%, respectively. 
On the other hand, RapidPatch incurred 7\% and 3.8\% overheads for Zephyr OS and FreeRTOS, respectively. 
%\karthik{Is this the same reason as we explain later, i.e., due to LLVM? If so, perhaps say it here as well.}
%This is because RapidPatch adds one trampoline for each function, but \SN{} adds multiple trampolines at different locations.
%For comparison, the popular UART printf function requires about 175,000 CPU cycles (about 2.7 $ms$ in nRF52840 board), 
%which is significantly slower than the software triggering method used in \SN{}.%~\cite{he2022rapidpatch}}.
%\mohsen{I used this sentence from RapidPatch but I run it on my board and write 175000 cycles instead of 
%180000 cycles, but should I still cite them?}

\subsubsection{Hotpatch Dispatching Overhead}
% \textcolor{red}{This subsection is done and the result is shown as a plot in Figure~\ref{dispatchoverhead}.
% This subsection, like the previous subsection, does not depend on CVE and can be measured separately.}
%Measure finding the correct patch among set of active patches 
%(like RapidPatch, use plot for different number of patches)
% \mohsen{I used the implementation of RapidPatch. Do we need to mention here or in implementation section?}
When a trampoline is triggered in a vulnerable function, \SN{} should select the appropriate patch
for that function from among the active patches - this is the hotpatch dispatching overhead.
Similar to RapidPatch, due to the memory limitation in embedded devices, we stored a maximum of 64 patches 
and evaluated the associated overhead. 
We evaluated this overhead by calculating the average time for different numbers of active patches 
(ranging from 1 to 64).
%\mohsen{Grammer issue.}
%the result is shown in Figure~\ref{dispatchoverhead}.
Our results indicate that the %the minimum and maximum 
overhead was 2.1 $\mu s$ for 1 patch and 4.4 $\mu s$ for 64 patches, respectively. 
This is because \SN{} uses binary search for finding the patch (similar to RapidPatch), and hence it takes $O(log n)$ time.
%\mohsen{Do you think this sentence "This is because \SN{} leverages binary search, and hence its time overhead is $O(log n)$." is necessary? If we remove it, it will free up a little space.} \karthik{We have space now, so I added it back.}
%\mohsen{Since we used the RapidPatch implementation, I didn't say their results for this part but in their paper, they have better result compare to our result!} 
%\karthik{We should still mention their result, and also mention we used a similar implementation as them.}
%\mohsen{Still I believe when they said using their implementation, we don't need to say that since they have weired result specifically for this! (1 to 2 microsecond!)}

%In this section, we assessed the patch dispatching overhead of \SN{}, 
%which involves selecting the appropriate patch from a set of active patches.
%To measure the time consumed by the patch dispatching function, we varied the number of active patches. 
%As many devices have memory constraints and can only install a limited number of patches, 
%we set the maximum active patch number to 64 in our experiments. 
%Our results, as illustrated in Figure~\ref{dispatchoverhead}, 
%show that the dispatching time increases with the number of patches, 
%taking approximately 4.4 $\mu$s for 64 patch points.

%\begin{figure}[ht]    
%	\centering
%	\includegraphics[scale=0.5]{Images/Dispatch_delay.png}
%	\caption{\SN{} Dispatch Overhead}
%	\label{dispatchoverhead}
%\end{figure}

\subsubsection{Hotpatch Execution Overhead}
\label{HED-sec}

\begin{table*}[!ht]
	\centering
	\caption{Execution time and memory overhead of \SN{} on STM32-F446RE, ESP-WROOM32, and STM32-L152RE boards 
	for different CVE types (Table~\ref{table-CVE-dataset}). Hotpatch execution overhead percentage compared to nRF52840 board are written in parentheses.
	A negative percentage value indicates a higher overhead compared to the nRF52840 board.}
	\resizebox{0.9\textwidth}{!}{%
	\begin{footnotesize}
	\begin{tabular}{c|c:c|c:c|c:c}
		& \multicolumn{2}{c|}{STM32-F446RE} & \multicolumn{2}{c|}{ESP-WROOM32} & \multicolumn{2}{c}{STM32-L152RE} \\
		\cline{2-3} \cline{4-5} \cline{6-7}
	CVE No.    &   \begin{tabular}[c]{@{}c@{}}Execution\\Time ($\mu s$)\end{tabular} & \begin{tabular}[c]{@{}c@{}}Memory\\Usage (Bytes)\end{tabular}  &   \begin{tabular}[c]{@{}c@{}}Execution\\Time ($\mu s$)\end{tabular} & \begin{tabular}[c]{@{}c@{}}Memory\\Usage (Bytes)\end{tabular} & \begin{tabular}[c]{@{}c@{}}Execution\\Time ($\mu s$)\end{tabular} & \begin{tabular}[c]{@{}c@{}}Memory\\Usage (Bytes)\end{tabular}  \\ 
	\hline \hline
	2020-10021  &   0.21  (32.2\%)        &   528          &   0.19 (38.7\%) & 732 &   0.75  (-141.9\%)        &   524 \\
	2020-10023  &   0.12  (40\%)          &   504         &   0.19 (5\%) & 576 &   0.37  (-85\%)        &   496 \\
	2020-10024  &   0.15  (42.3\%)        &   392          &   0.14 (46.1\%) & 524 &   0.5  (-92.3\%)        &   384 \\
	2020-10028  &   0.23  (41\%)          &   400         &   0.26  (33.3\%) & 556 &   0.78  (-100\%)        &   392 \\
	2020-10062  &   0.33  (31.2\%)         &  780          &   0.33  (31.2\%) & 1280 &  1.03  (-114.5\%)        &   764 \\
	2020-10063  &   0.21  (41.6\%)          & 400          &   0.16  (55.5\%) & 732 &   0.81  (-125\%)        &   392 \\ 
	\hline
	2018-16524  &   0.25  (32.4\%)           &  548        &   0.31 (16.2\%) & 720 &   0.84  (-127\%)        &   396 \\
	2018-16528  &   0.24 + 0.27 (33.7\%)      & 468 + 476        &   0.37 + 0.35 (6.5\%) & 708 + 772 &   0.71 + 0.84  (-101.3\%)        &   460 + 469  \\
	2018-16603  &   0.27  (34.1\%)             & 516      &   0.35  (14.6\%) & 744 &  0.96  (-134.1\%)        &   512 \\ 
	\hline 
	2017-2784   &   0.18  (30.7\%)              & 380      &   0.17 (19\%) & 488 &   0.65  (-150\%)        &   372 \\
	2020-17443  &   0.21  (32.2\%)              &  476     &   0.18 (41.9\%) & 648 &   0.75  (-141.9\%)        &   468 \\
	2020-17445  &   0.27  (34.1\%)               &  560    &   0.29 (29.2\%) & 864 &   0.9  (-119.5\%)        &   552 \\                                            
		\hline
	\textbf{Average} & \textbf{0.24 (35.1\%)} & \textbf{535} & \textbf{0.27 (27.3\%)}  & \textbf{778.5} & \textbf{0.82 (-121.6\%)}  & \textbf{515}
\end{tabular}
\end{footnotesize}

}
		\label{table-CVE-dataset-appendix}
\end{table*}
To determine the execution time of each hotpatch, we measured the times by directly executing each patch 
multiple times
on the nRF52840, under different input values (similar to RapidPatch) and computed the average time. %\karthik{How did you get the input values?}
The execution time for each patch is summarized in Table~\ref{table-CVE-dataset} (column \textit{Execution Time}), and ranges from 0.20 to 0.77 $\mu$s, with an average value of $0.37\mu$s. %\karthik{Please check.} 
Thus, all patch execution times were less than 1 $\mu s$, indicating their negligible overheads.
%Furthermore, it is important to acknowledge that this overhead depends 
However, the precise overhead depends on the specific implementation details of the official patch
employed by RTOS developers, such as the number of instructions in the patch. %or the coding approach employed by the RTOS developer. %\karthik{What's the corresponding number for RapidPatch?} 

Furthermore, we evaluated \SN{} on the remaining CVEs other than those in Table~\ref{table-CVE-dataset}. As mentioned in Section~\ref{sec-eval-effectiveness}, 
\SN{} could not generate hotpatches for six of the 62 CVEs. Another six of the CVEs did not have official patches available, and hence we excluded these CVEs as well (RapidPatch has also excluded these six CVEs in their evaluation). 
% \karthik{Please check}
%\karthik{Is this correct? Can we say whether RapidPatch included them?}
%\mohsen{Sorry, I don't understand the comment. Rapidpatch appears not to have thoroughly tested any of the 50 CVEs, as we stated in the paper. They only created a private GitHub repository that collects the descriptions of each CVE. For six of these CVEs, they indicated that there is no patch available, which is why I stated the same. Perhaps there is a patch available, but I couldn’t find one. Should we consider changing the term?}
For the remaining 38 CVEs, (note that we already evaluated 12 of the 50 CVEs earlier)
%\mohsen{Is it ok we said we evaluated AutoPatch on 12 CVEs in the beginning of section 6.2.}
we found that the average execution overhead of the generated hotpatches is approximately 0.33 $\mu s$. 
The average memory overhead for these CVEs is about 400 bytes. 
%Note that seven of these 62 CVEs have not yet been officially patched. 
%Also, 
These overheads are comparable to those incurred by \SN{} on the original 12 CVEs in Table~\ref{table-CVE-dataset}.
The detailed results and information for the remaining 50 CVEs are shown in Table~\ref{tab:CVEdatasetextra} in Appendix~\ref{sec:appendix-dataset}.

\subsubsection{Total Patch Overhead}
\label{subsubsec-totaldelay}

%
%% ***** ATTENTION ATTENTION: I wrote LIKE RAPIDPATCH, also What should I do for maximum time when the triggering is in the loops?
%
We evaluated the entire patch execution time overhead, which encompasses the three distinct overheads: 
trampoline execution, patch dispatching, and hotpatch execution by using the CVEs listed in Table~\ref{table-CVE-dataset}.
%\karthik{This is isn't the patching process overhead, but the patch execution overhead!}
The results indicate that, on average, the patch execution takes less than 12.7 $\mu s$ (3.3 $\sim$ 12.7 $\mu s$), which is negligible for most practical purposes.

We also measured the memory overhead of storing the executable file of each generated hotpatch for each CVE, 
as shown in 
the \textit{Memory Usage} column of Table~\ref{table-CVE-dataset}. %The results indicate 
We see that \SN{} 
incurs an average memory overhead of about 535 bytes on the device. %, which is negligible.
%\karthik{I moved this here as this is incurred on the device at runtime.} %\karthik{Again, we should compare with RapidPatch's overhead here}
%Furthermore, the overhead of trampoline execution and patch dispatching are constant, 
%while the time complexity of implementing the official patch directly influences this overhead.

%\TC{Similar to Section~\ref{TED-sec}, we evaluated the total patching delay incurred by \SN{} on different
%real applications provided by ZephyrOS, and the results show 0.34\% to 3.9\% overhead.} \karthik{I'd expect this to be the case as the overhead is dominated by the instrumentation overhead. Perhaps mention this here.}

\subsubsection{Generality}
\label{sec:evaluation-extra}

%\mohsen{Should be as a new subsection or it is good but should I change the title "Efficiency"? Should I report all CVE results for all boards? (it takes a lot of space, I think)}
As discussed earlier, since \SN{} works on LLVM IR to generate hotpatches, it is a platform- and architecture-independent method.
%To evaluate this, 
%we choose two other different boards with various architectures and specifications (Table~\ref{tab:specification}) 
%to generate hotpatches for 12 CVEs mentioned in Table~\ref{table-CVE-dataset}.
We have successfully executed the hotpatches generated by \SN{} on the STM32-F446RE\footnote{We set the board frequency to 96 MHz.}, 
STM32-L152RE, and ESP-WROOM32 boards (Table~\ref{tab:specification}),  
%, which have different architectures and specifications (Table~\ref{tab:specification})
without any modifications to \SN{}'s implementation. 
%\karthik{If you didn't modify the implementation, then why did have to you port it?}
The results are shown in Table~\ref{table-CVE-dataset-appendix}, including hotpatch execution and memory overhead. They  
demonstrate \SN{}'s ability to generate hotpatches for the mentioned CVEs in Table~\ref{table-CVE-dataset} 
across these boards without implementation changes. 
The results show that, on average, the hotpatch execution overhead is
%for both boards  
0.24 $\mu s$ (0.12 $\sim$ 0.51 $\mu s$), 0.27 $\mu s$ (0.14 $\sim$ 0.72 $\mu s$)
and 0.82 $\mu s$ (0.37 $\sim$ 1.55 $\mu s$)
for the STM32-F446RE, ESP-WROOM32 and STM32-L152RE boards, respectively. 
% These
% \textcolor{blue}{The STM32-F446RE and ESP-WROOM32 incur lower overheads compared to the}
% are lower than the overhead for the 
% nRF52840 board 
% as these boards have higher frequency processors (Table~\ref{tab:specification}). 
%as microcontroller units (MCUs) with higher frequencies have smaller overhead~\cite{he2022rapidpatch}. 
%\karthik{I added this. Please check}
% Detailed results are in Section~\ref{sec:evaluation-extra}.
%\mohsen{You delete the summary results from here. We don't need to report those in the main text?}
% \textcolor{blue}{On the other hand, the STM32-L152RE board incurs a higher overhead 
% compared to the nRF52840 board due to its lower processor frequency (32 MHz).}
% \karthik{The above text just repeats the first point in the paragraph below. Perhaps merge it with that.}

% \mohsen{TODO: What about the new board. Add when the results are ready.}
From the results in  Table~\ref{table-CVE-dataset-appendix}, three main observations can be made. 
First, performance overheads are lower for the first two boards 
(i.e., STM32-F446RE and ESP-WROOM32) due to their higher frequencies compared to nRF52840 (Table~\ref{tab:specification}). 
On the other hand, the STM32-L152RE board incurs a higher overhead 
compared to the nRF52840 board due to its lower processor frequency (32 MHz).
The overhead difference of each CVE between each board and nRF52840 is written in parentheses.
Hence, a higher performance overhead indicates a lesser improvement compared to the nRF52840 board.

Second, the overhead pattern of CVEs is nearly identical between the nRF52840 and STM32-F446RE boards, 
which have the same architecture but differ in specifications. 
For instance, CVE-2020-10023 and CVE-2018-16528 incur the minimum and maximum overheads, respectively.
In contrast, the overhead pattern differs for the ESP-WROOM32 board, 
which has different architecture and specifications. 
For example, CVE-2020-10024 has the lowest overhead but similarly CVE-2018-16528 has the highest overhead.

Third, even though the ESP-WROOM32 board operates at a higher frequency than the STM32-F446RE, the overhead of the ESP-WROOM32 is lower than that of the STM32-F446RE in some 
CVEs, while in other CVEs, it is higher.
This is because the overhead is not affected by frequency alone; various factors, including architecture, contribute to this metric.

\subsubsection{Comparison with RapidPatch}
\label{subsubsec-comparerapid}
%In order to assess the performance of our proposed approach, we conducted a comparative evaluation with RapidPatch, 
%the state-of-the-art hotpatching technique.
%To evaluate their technique in terms of patch execution time and memory usage, 
RapidPatch used the same nRF52840 board and the same CVE dataset for evaluating their memory and execution time overheads, as well as the same RTOSes. %\karthik{Please check}
Therefore, we directly compare our results with their results. 
Note that RapidPatch only reported results for the 12 CVEs listed in Table~\ref{table-CVE-dataset} on the nRF52840 board, but not on the other boards.
\begin{figure*}[!ht]
	\centering
	\begin{subfigure}{0.45\textwidth}
		\centering
		\includegraphics[width=\textwidth]{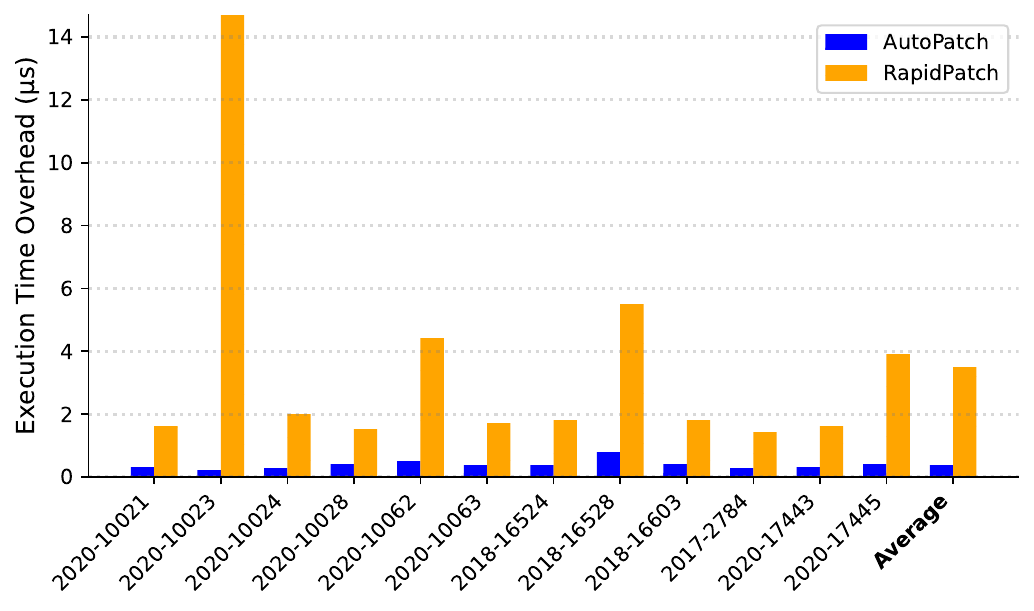}
		\caption{Hotpatch Execution Time Overhead}
		\label{fig:ecompare}
	\end{subfigure}
	\hfill % Add horizontal space between the subfigures
	\begin{subfigure}{0.46\textwidth}
		\centering
		\includegraphics[width=\textwidth]{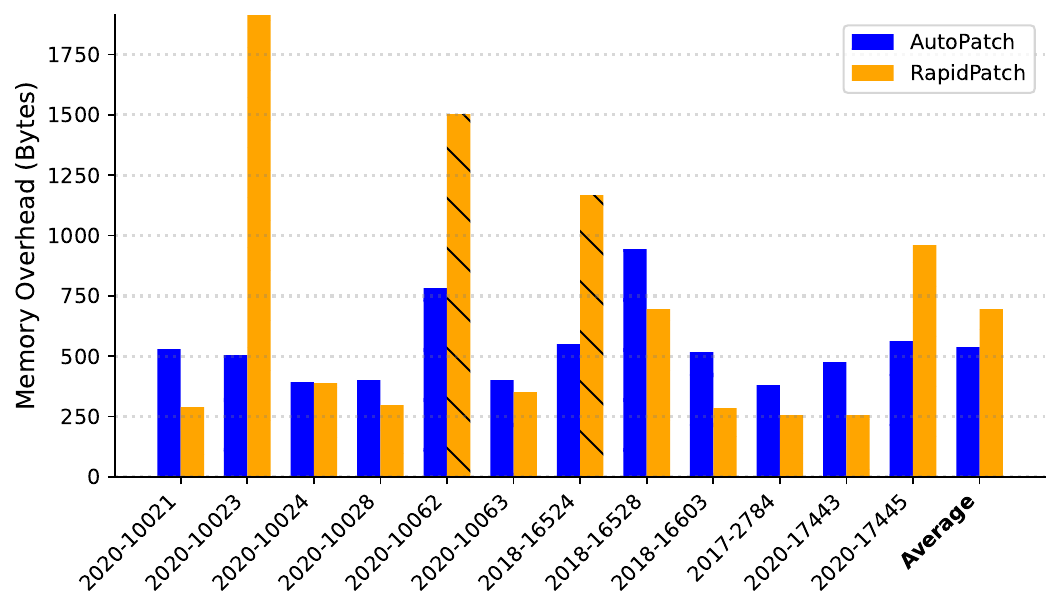}
		\caption{Memory Overhead}
		\label{fig:mcompare}
	\end{subfigure}
%\vspace{-3mm}
	\caption{Comparison of execution time and memory overheads between \SN{} and RapidPatch~\cite{he2022rapidpatch} for each CVE.}
	%\vspace{-5mm}
	%\caption{Comparison of Overheads between \SN{} and RapidPatch~\cite{he2022rapidpatch} for each CVE. (a) Execution time, and (b) Memory.}
	\label{fig:compare}
\end{figure*}

Figure~\ref{fig:ecompare} and 
Figure~\ref{fig:mcompare} present the execution time %overhead 
and memory overhead results, respectively, for each CVE, comparing \SN{} and RapidPatch. 
%\mohsen{I commented the sentence "we are slightly higher than them".}
%As shown in Figure~\ref{fig:mcompare}, \SN{} incurs a slightly higher memory overhead than RapidPatch. 
On average, RapidPatch needs 695 bytes (256 $\sim$ 1914 bytes), 
while \SN{} needs 535 bytes (380 $\sim$ 944 bytes),
resulting in a 23\% reduction compared to RapidPatch.
Note that, upon reviewing the RapidPatch code, we identified that their reported memory overhead 
only accounts for the locally copied patch code, 
but not the storage for the generated hotpatch file on the device. We included this overhead, and hence there is a slight variance between the memory overhead  in Figure~\ref{fig:mcompare} and that in their paper.
%It is important to note that since \SN{} works on LLVM, 
%\mohsen{I commented the sentence "the reason why we are higher than them." because we are not anymore!}
%This is because \SN{} uses LLVM, which 
%adds additional information to the file for compiling. However, this overhead can be reduced by optimizing the executable file. 

%\karthik{Added this.}  
%and with increasing optimization level, the generated file may have less memory than these values without any optimization.
%From RapidPatch:
%Overall, \SN{} exhibits a slightly higher, yet comparable, memory overhead when compared to RapidPatch.
For the execution time overhead, 
we compared \SN{}'s time with the Just-In-Time (JIT) mode of RapidPatch, 
which is faster than the interpreter mode of RapidPatch~\cite{he2022rapidpatch}.
%the fastest mode and represents the best results of RapidPatch in the Interpreter mode comparison. 
Therefore, our comparison is conservative. 
We find that on average, RapidPatch requires 4.53 $\sim$ 25.6 $\mu s$ %in JIT mode, %and it reaches about 140 $\mu s$ 
%in interpreter mode
while \SN{} needs 3.3 $\sim$ 12.7 $\mu s$ to fix the security vulnerability. 
Therefore, \SN{} is about 50\% faster than RapidPatch on average for the total patch execution time, 
even when RapidPatch is executed in the faster JIT mode. 
% Furthermore, \SN{} achieves a notable 89\% improvement, 
% with an average execution time of 0.37 $\mu s$ for executing the hotpatch 
% compared to RapidPatch's average of 3.49 $\mu s$ in JIT mode (Figure~\ref{fig:ecompare}).
Furthermore, \SN{} %achieves a notable 89\% improvement, 
has an average execution time of 0.37 $\mu s$ for executing the hotpatch 
compared to RapidPatch's average of 3.49 $\mu s$ in JIT mode (Figure~\ref{fig:ecompare}). This is a speedup of about $9.5$x over RapidPatch.
% \karthik{Actually, we should measure the speedup, rather than execution time here. Please check.}

%\karthik{As we discussed, we can perhaps dilute this a little bit and instead say we're comparable to RapidPatch despite needing no manual effort.}
%\mohsen{Added to Intro section (contribution part).}

\subsubsection{Comparison with HERA}
We did not have access to the HERA's implementation~\cite{niesler2021hera}, as their source code was not publicly available. Therefore, for HERA's overhead,  
we used the numbers provided by RapidPatch. They  
evaluated HERA using CVE-2018-16601 on STM32L475 board at 80 MHz - this is different from the nRF52840 board, which operates at 64 MHz. 
As a result, we estimated HERA's overhead for the nRF52840 board, 
%The estimated overhead for running CVE-2018-16601 with HERA on the nRF52840 board is 
as approximately 0.35 $\mu s$. 
On the same board, the overhead incurred by \SN{} is around 0.45 $\mu s$. 
HERA incurs a slightly lower overhead due to its hardware-based approach, and as it does not add any additional instructions.
%which only requires executing patch instructions. 
In contrast, \SN{} adds instructions to the patch to make it executable by 
 the trampolines. However, \SN{}'s overhead is only slightly higher than HERA's. Note that RapidPatch's overhead for CVE-2018-16601 on the nRF52840 board in JIT mode is approximately 1.9 $\mu s$, which is  higher than both \SN{} and HERA.

%In summary, \SN{} has a slightly higher but comparable hotpatch execution overhead compared to HERA. 
As previously discussed in Sections~\ref{sec-intro} and \ref{sec-otherwork}, HERA cannot be used with all devices. This is because 
this hardware feature (i.e., FPB) 
%is not supported by all devices and 
exists only in ARM Cortex-M3 and M4 processors. 

\label{evaluatesec}

% \documentclass[../main.tex]{subfiles}
% \begin{document}

\section{Related Work}
%This section categorizes prior work on hotpatching frameworks into two classes: 
%(1) traditional and Android systems update,  and (2) embedded devices firmware update.

\textbf{Traditional and Android Systems Update.}
%Researchers have been working on hotpatch ideas for many years,  
%,and their goal has been to provide this kind of approach 
%for traditional software. 
Segal et al.~\cite{segal1993fly,frieder1991dynamically},
introduced a PODUS system 
(the procedure oriented dynamic updating system) for updating C applications.
Other work has focused on server applications 
%and as a result, approaches have been proposed to solve various challenges 
in this field~\cite{altekar2005opus,chen2007polus,payer2013hot,giuffrida2016automating}.
Some of these approaches %are based on the compiler and 
change the compiler~\cite{altekar2005opus,chen2007polus}, 
while other approaches change the programming language~\cite{smith2012towards}. 
The main challenge of hotpatching for these systems is that after making changes and rebooting the device, 
the system state must also be maintained. 
%In other words, since the systems or programs need to be restarted after the changes, 
%the system state is erased, so the current state must be saved so that after the restart, 
%the system starts working from the saved state.

Hayden et al.~\cite{hayden2012kitsune} proposed Kitsune that leverages update points 
where the states of the program are easily accessible, and these points are specified by the programmer.
Guiffrida et al.~\cite{giuffrida2013back} used fault-tolerant state transfer method for solving this challenge, 
while Makris et al.~\cite{makris2009immediate} leveraged an algorithm called stack reconstruction.
%In the past, traditional systems relied on these hotpatching methods to perform dynamic updates, 
%which involved relocating executable files.
%By increasing the number of embedded devices, hotpatching has gained attention 
%to fix security vulnerabilities.
%But, the new hotpatching approach must take into consideration the memory and real-time limitations
However, the unique characteristics of real-time systems, such as timeliness, resource limitations, 
and the necessity for predictability, limit the applicability of these techniques.

% However, these systems are not suitable for the memory and real-time limitations 
% of real-time embedded devices, as they were not designed for the same.\par 
% \karthik{What specifically do they do that makes them unsuitable?}

Ksplice~\cite{arnold2009ksplice} proposed a generic hotpatching technique for Linux systems that operates  at the object code level. 
Ksplice identifies the differences between the vulnerable kernel and the patched kernel, 
locates the modified functions, and adds the entire function along with the patch to the device at runtime. 
However, since Ksplice stores the complete patched function beside the vulnerable function,
it requires a large amount of memory. 
In contrast, \SN{} does not need to store the entire patched function and it only stores
the generated hotpatch. This makes it better suited for memory-constrained devices such as real-time embedded devices.
%\karthik{Why is this important?}
%\mohsen{I said this because of Reviewer C. Delete it?}
Also, Ksplice needs to modify the code at execution time to apply the patch, which is not suitable for embedded devices due to the hardware characteristics, as discussed in Section~\ref{hotpatchback}.
%\karthik{Why not? I don't understand.}
%This approach identifies the differences between the vulnerable kernel and the patched kernel, 
%locates the modified functions, and adds the entire function along with the patch to the device at runtime. 
%However, because the complete function is stored and executed instead of just the patch instructions, 
%this method requires a large amount of memory. 
%Additionally, since Ksplice needs to modify the code at execution time to apply the patch, 
%it is not suitable for use in embedded devices. 

The authors in~\cite{chen2018instaguard,chen2017adaptive} proposed two hotpatching techniques for Android
devices by writing hotpatches in new languages (i.e., GuardSpec and Lua).
However, rewriting a patch in a new language can be a time-consuming and error-prone process.
Furthermore, KARMA~\cite{chen2017adaptive} requires code rewriting to insert trampolines, 
which needs to reboot the embedded devices (Section~\ref{hotpatchback}) 
while Instaguard~\cite{chen2018instaguard} relies on hardware techniques to trigger the patch, 
which are not commonly supported in embedded devices.

Zhang et al.~\cite{zhang2017embroidery} proposed Embroidery, a patching system designed
to fix vulnerabilities on outdated Android devices by using binary code. 
However, Embroidery requires rewriting the kernel and modifying the kernel memory region, 
which results in device reboots and violates one of our main requirements (Section~\ref{sec-challenge}).

Finally, Bowknots~\cite{talebi2021undo} %for kernel bugs 
attempts to maintain system functionality and nullify the side effects of currently executing 
system calls when a kernel bug is triggered, and no official patch is available. 
This method assumes that the official patch has not yet been released, 
focusing on the window between the discovery of the vulnerability and the release of the patch.
Therefore, our work is orthogonal to this line of work.

\textbf{Embedded Devices Update.}
%In recent years, with the increasing use of embedded %importance and number of embedded devices and their 
%use in critical applications 
%such as cyber physical systems and power grid, 
In recent years, the patching and updating of embedded devices firmware  has become 
important~\cite{park2010dynamic,arakadakis2021firmware,felser2007dynamic,dong2013r3}.
Most previous approaches used the over-the-air (OTA) method to update the system and 
fix bugs and vulnerabilities in the program~\cite{bauwens2020over,arakadakis2021firmware}.
There are challenges in the OTA method, such as secure updating and reduction of transmitted data,
which researchers~\cite{ammar2020verify,panta2009zephyr,bauwens2020over} have addressed. 
These OTA methods require reprogramming, and since embedded devices run code on the flash-based ROM, 
they require that the entire flash sector be erased. 
Thus, the system must reboot, which halts the running task, making them unsuitable for mission-critical applications, which is our focus. \par

% ***I commented these two approaches since "we don't have enough space!***%
% Furthermore, there are other approaches~\cite{felser2007dynamic,park2010dynamic} for embedded devices 
% that used new architecture to update the devices.
% The authors in~\cite{felser2007dynamic,park2010dynamic} proposed an architecture 
% for patching sensor nodes and gathering and distributing updates, respectively.\par
% \karthik{What's the drawback of these approaches?}

% Over time, real-time embedded devices played a significant role in important parts of human life, 
% \karthik{Why are you telling me this now - you already mentioned this many times before!}
% *****......... and these methods could not be used for such devices 
% due to many challenges~\cite{niesler2021hera,wahler2009dynamic,he2022rapidpatch,wahler2014disruption}.
ChkUp~\cite{wuyour} proposed a technique to detect vulnerabilities in the firmware update process by using cross-language inter-process
control flow analysis and program slicing.
While there have been hotpatching approaches that do not require a reboot for embedded devices~\cite{niesler2021hera,wahler2009dynamic,he2022rapidpatch,wahler2014disruption}, each of the approaches has problems that make them unsuitable for our purposes (see Section~\ref{sec-otherwork} for details). %\karthik{I suggest just pointing the reader to the previous discussion in Section 3 instead of repeating it here.}
%HERA~\cite{niesler2021hera}, the first hotpatching approach for real-time embedded devices, 
%use Flash Patch and Breakponit Unit (FPB) to fix security vulnerabilities. 
%Unfortunately, this FPB feature is only available on ARM Cortex M3/M4 devices 
%and as a result this approach is not applicable to other devices.
%To solve this problem, RapidPatch~\cite{he2022rapidpatch} was proposed, which uses the hardware method (i.e., FPB and KProbe) 
%and in its absence, the software method to trigger the patch.
%Furthermore, their goal was to provide a hotpatch for different devices with different hardware architectures 
%that use the same RTOS with similar vulnerability.
%These devices require very low overhead approaches while Rapidpatch requires VM on the device. 
%Also, after programming the patch in C language, the RTOS developer has to rewrite the patch in eBPF language, 
%which is a time consuming and error-prone process. \SN{} enables an automatic method to generate the hotpatch for
%real-time embedded devices which does not require another language and VM. 
%\SN{} also uses a software approach to trigger the patch.
%Add papers that they cite to RapidPatch and HERA here:
%OPEC \cite{zhou2022opec}
%Perils and Mitigation of Security Risks of Cooperation in Mobile-as-a-Gateway IoT \cite{zhou2022perils}
%VERI \cite{cheng2023veri}
%A summary of state-of-the-art hotpatching methods %for various types of devices, 
%Table~\ref{tab:compare} in the Appendix summarizes them.
\par

\textbf{Summary.}
Table~\ref{tab:compare} provides a summary of the state-of-the-art hotpatching methods across different types of devices, 
including Android, real-time embedded systems, and general-purpose devices. 
As we analyze each of the proposed 
approaches~\cite{niesler2021hera,he2022rapidpatch,chen2018instaguard,chen2017adaptive,arnold2009ksplice}, 
we identify limitations that render them unsuitable for use on embedded devices. 
In brief, the identified limitations of these hotpatching methods
include hardware dependencies (e.g., FPB) for patch triggering, 
dependence on an additional programming language and VM to execute the patch, 
and runtime modifications of the program code, all of which render them impractical for use on 
real-time embedded devices.
To address these limitations and automatically generate patches for device vulnerabilities, 
we introduce \SN{}.

\definecolor{Gray}{gray}{0.9}
\newcolumntype{g}{>{\columncolor{Gray}}c}
\begin{table}[!ht]
  
  \begin{center}
   \caption{A subjective comparison between \SN{} and state-of-the-art hotpatching methods proposed
    for Android (A) devices, real-time Embedded (E) devices, and General Purpose (GP) devices. CCR and RP denote change code at runtime
    and rewrite patch in other language, respectively.}
   \label{tab:compare}
   \begin{tabular}{c|ccccc|g}
    Objectives & {\rotatebox[origin=c]{90}{ HERA \cite{niesler2021hera}}} & {\rotatebox[origin=c]{90}{RapidPatch \cite{he2022rapidpatch}}} & {\rotatebox[origin=c]{90}{ Instaguard \cite{chen2018instaguard}}}  & {\rotatebox[origin=c]{90}{ KARMA\cite{chen2017adaptive}}} & {\rotatebox[origin=c]{90}{ Ksplice \cite{arnold2009ksplice}}} & {\rotatebox[origin=c]{90}{ \SN{} }} \\ 
    \hline \hline
    Target& E& E & A & A & GP & E\\ 
    w/o Need Hardware& \xmark & \cmark & \xmark & \cmark & \cmark & \cmark\\
    w/o CCR& \cmark & \cmark & \cmark & \xmark & \xmark  & \cmark\\ 
    w/o Need VM& \cmark & \xmark & \cmark & \xmark & \cmark  & \cmark\\ 
    w/o RP& \cmark & \xmark & \xmark & \xmark & \cmark & \cmark\\ 
    Common Patch& \xmark & \cmark & \xmark & \cmark & \cmark  & \cmark\\ 

    \hline 
   \end{tabular}
  \end{center}
 \end{table}

%\subsection{Summary}
%Table~\ref{tab:compare} introduces a subjective comparison of state-of-the-art hotpatching methods on 
%different types of devices (i.e., android, real-time embedded and general purpose devices) and \SN{}.

%As can be seen, each of the proposed
%approaches~\cite{niesler2021hera,he2022rapidpatch,chen2018instaguard,chen2017adaptive,arnold2009ksplice} 
%has problems that make it unusable for embedded devices. As a result, we introduced \SN{}{} 
%to fix previous work problems as well as automatically generate patches to fix device vulnerabilities.

% \end{document}

\label{releatedsec}

% \documentclass[../main.tex]{subfiles}
% \graphicspath{{\subfix{../images/}}}
% \begin{document}

\section{Discussion}
\label{sec:discussion}

%Like RapidPatch I said "common hotpatching problems":
%But some of them are just for me (using one board and inserts trampoline in different places.)
Although \SN{} has addressed hotpatching challenges, it still has three main limitations. 
%In this section, we explore these limitations and provide insights that can guide future research endeavors.
First, while the evaluation in Section~\ref{TED-sec} demonstrated 
that the software triggering approach does not impose significant overhead on the device and application, 
it is possible to further minimize this overhead by reducing the number of trampoline locations. 
Previous work~\cite{he2022rapidpatch} has shown that the majority of vulnerabilities are associated 
with network payload processing functions, and therefore the instrumentation 
can be done only for packet ingress functions. 
Additionally, given that embedded devices are designed for specific purposes, 
they consistently use certain OS functions and libraries throughout their operational lifespan. 
Consequently, instrumenting only those specific functions can effectively decrease 
the overhead imposed on the device. These changes will not require modifications in the IC and AC.

Second, similar to prior work, we made the assumption that the official patches released by RTOS developers 
for resolving vulnerabilities are free of bugs. 
However, %it is feasible to incorporate an additional 
%we can incorporate a component 
%that assesses official patches for potential bugs and ensures their correctness before generating the hotpatch.
%\mohsen{I think it needs citation!Any suggestion for that?}
there are many formal verification~\cite{jacobs2011verifast} techniques that can be used to verify the patches.
%\mohsen{Do you think is it good and enough? Or should we also mention here for correctness of generated hotpatches?}
%\karthik{Add citation}
%\mohsen{Also for after generating the patch!}

Third, similar to RapidPatch~\cite{he2022rapidpatch} and HERA~\cite{niesler2021hera}, \SN{} does not support official patches 
that require changes to either data structures or macro definitions (Section~\ref{sec-intro}). However, we found that only 2 out of 62 CVEs had official patches that made such changes.

\label{discussionsec}

% \documentclass[../main.tex]{subfiles}
% \graphicspath{{\subfix{../images/}}}
% \begin{document}

\section{Conclusion}

% In this article, we present AutoPatch, the first automatic hotpatching framework for real-time embedded devices. 
% AutoPatch leverages a novel combination of static instrumentation and analyzing to automatically 
% generate semantic preserving hotpatch based on the official patch, 
% allowing RTOS developers to generate hotpatches without manual effort. 
% We evaluated the effectiveness and efficiency of \SN{} on different major RTOSes with various real CVEs. 
% Our technique correctly generate all the hotpatches by preserving their semantics with negligible 
% runtime overhead and memory.

We propose \SN{}\footnote{We have made our code and data publicly available at \url{https://github.com/DependableSystemsLab/AutoPatch}.}, the first automatic hotpatching framework for real-time embedded devices. 
\SN{} addresses challenges existing in current hotpatching techniques by combining static instrumentation 
and analysis techniques to \emph{automatically} generate functionality-preserving hotpatches based on official patches.
%Unlike existing hotpatching techniques, \SN{} %addresses challenges existing in current hotpatching techniques by 
%combines static analysis techniques and instrumentation, to \emph{automatically }generate functionality-preserving hotpatches based on official patches.
\SN{} is implemented using the LLVM compiler and is hence portable to different platforms. 
%In our software triggering technique, we determined four instrumentation locations, which are necessary
%to avoid the use of estimation and optimization techniques in static analysis.
%Also, our comprehensive investigation of various CVEs in popular RTOSes has shown
%these locations are adequate.
We evaluated \SN{} using real CVEs on four different embedded boards with various architectures  
running different RTOSes. Our evaluation found that \SN{} 
successfully generated the hotpatches while preserving their \RE{} for over 90\% of CVEs. 
Further, \SN{} had negligible runtime overhead and minimal memory consumption, 
and incurred lower performance and memory overheads than a prior state-of-the-art hotpatching approach, despite needing no manual effort.  %\textcolor{red}{and also incurred lower memory overhead on average}. 
Finally, \SN{} generated hotpatches can be executed on different boards, without any changes, thus demonstrating its generality.
%, but had slightly higher memory overhead. %\karthik{Added this}

% In this article, we propose \SN{}, the first automatic hotpatching framework for real-time embedded devices.
% \SN{} utilizes a novel combination of static instrumentation and analysis techniques to automatically generate 
% semantic-preserving hotpatches based on official patch, thereby enabling RTOS developers to generate hotpatches without manual intervention.
% We also conducted a comprehensive investigation of various CVEs in popular RTOSes 
% and identified instrumentation locations that are crucial for preventing the utilization 
% of estimation or optimization techniques in hotpatching.
% %We conducted a comprehensive evaluation of \SN{} on major RTOSes, 
% %considering a diverse set of real CVEs. 
% Our evaluation demonstrated the effectiveness and efficiency of \SN{}, 
% as it successfully generated all required hotpatches while preserving their semantics. 
% Moreover, \SN{} exhibited negligible runtime overhead and memory consumption.
%We evaluated the effectiveness and efficiency of \SN{} on different major RTOSes with various real CVEs.
%Our technique correctly generate all the hotpatches by preserving their semantics with negligible runtime
%overhead and memory.

% \end{document}
\label{conclusionsec}

\section*{Acknowledgements}
This work was partially supported by the Natural Sciences
and Engineering Research Council of Canada (NSERC), the Department of National Defense (DND), Canada, and
a Four Year Fellowship from UBC. 
We also thank the anonymous reviewers of CCS 2024. % for their helpful comments.

%\input{Sections/Motivation}
%\input{Sections/Design}
%\input{Sections/Results}
%\input{Sections/Conclusion}

\bibliographystyle{ACM-Reference-Format}
\balance
\bibliography{main}

%-------------------------------------------------------------------------------
\appendix
%-------------------------------------------------------------------------------

%\input{Sections/ICAlgorithmA.tex}

%\input{Sections/ACAlgorithmA.tex}

% \input{Sections/ImplementationA.tex}

% \input{Sections/ResultsSA.tex}
% \input{Sections/RelatedA.tex}

\section{LLVM IR Example}
%\karthik{If you add it, you need to explain it using line numbers. What's the takeaway here?}
\begin{figure}[H]
    \lstset{basicstyle=\scriptsize}
\begin{lstlisting}[caption=The LLVM IR hotpatch generated by \SN{} for the first part of CVE-2020-10062 (Listing~\ref{list5ip}).,label=llvmex]
1 %stack_frame = type { i32, i32, i32, i32, i32, i32, i32, i32 }
2 define i64 @filter_10062_1(%stack_frame* %0) {
3 entry:
4  %op = alloca i32, align 4
5  %ret_code = alloca i32, align 4
6  %1 = getelementptr inbounds %stack_frame, %stack_frame* %0, i32 0, i32 0
7  %r0 = load i32, i32* %1, align 4
8  %2 = getelementptr inbounds %stack_frame, %stack_frame* %0, i32 0, i32 1
9  %r1 = load i32, i32* %2, align 4
10  %3 = getelementptr inbounds %stack_frame, %stack_frame* %0, i32 0, i32 2
11  %r2 = load i32, i32* %3, align 4
12  %4 = getelementptr inbounds %stack_frame, %stack_frame* %0, i32 0, i32 2
13  %r3 = load i32, i32* %4, align 4
14  %5 = getelementptr inbounds %stack_frame, %stack_frame* %0, i32 0, i32 6
15  %r6 = load i32, i32* %5, align 4
16  %6 = getelementptr inbounds %stack_frame, %stack_frame* %0, i32 0, i32 7
17  %r7 = load i32, i32* %6, align 4
18  store i32 0, i32* %op, align 4
19  store i32 0, i32* %ret_code, align 4
20  %bytes = alloca i8, align 1
21  %7 = trunc i32 %r0 to i8
22  store i8 %7, i8* %bytes, align 1
23  br label %do.body
24 do.body:             ; preds = %entry
25  %8 = load i8, i8* %bytes, align 1, !dbg !0
26  %9 = zext i8 %8 to i32, !dbg !0
27  %10 = icmp sge i32 %9, 4, !dbg !24
28  br i1 %10, label %if_then, label %return
29 if_then:             ; preds = %do.body
30  store i32 1, i32* %op, align 4
31  store i32 -22, i32* %ret_code, align 4
32  br label %return
33 return:               ; preds = %do.body, %if_then
34  %op_val = load i32, i32* %op, align 4
35  %11 = zext i32 %op_val to i64
36  %ret_code_val = load i32, i32* %ret_code, align 4
37  %12 = zext i32 %ret_code_val to i64
38  %shifted_op = shl i64 %11, 32
39  %op_ret_sum = add i64 %shifted_op, %12
40  ret i64 %op_ret_sum
41 }
\end{lstlisting}
\end{figure}

Listing~\ref{llvmex} illustrates the generated hotpatch in LLVM IR format for CVE-2020-10062 by \SN{}.
For the sake of simplicity, 
we explain only the first part of CVE-2020-10062 (i.e., lines 6 to 7 in Listing~\ref{list5ip}).
%\mohsen{I had to remove part of the patch from this code, it was too big.}
As discussed in Section~\ref{patchasec}, after selecting the best trampoline and analyzing the official
patch by AC, AC stores the obtained instructions in a template function (e.g., $filter\_10062\_1$) 
that takes a data structure as an argument (i.e., $stack\_frame$).

As can be seen in the Listing~\ref{llvmex}, 
lines between 4 and 19 are for initialization, and in lines 20 to 22, 
the patch's variable value (i.e., $bytes$) is sent to the hotpatch from the vulnerable function through the $stack\_frame$.
Then, AC inserts the obtained instructions from analysis phase (lines 25 to 28) in appropriate basic blocks, if they consist of several basic blocks.
In CVE-2020-10062, since the best trampoline is close to the official patch and there is no change in the patch variables between the trampoline and the official patch, it consists of a basic block (i.e., $do.body$).  
Finally, based on the result of the comparison (line 28), it is either entered into the $if\_then$ basic block and the values of $op$ and $ret\_code$ 
are changed accordingly, so that \SN{} can perform one of the two operations, DROP or REDIRECT (see Section~\ref{patchisec}), 
which is equivalent to \textit{\(\text{"return -EINVAL"}\)} in the official patch;
or it is entered into $return$ basic block, which the values of $op$ and $ret\_code$ remain 0 (i.e., PASS), 
and it means $bytes$ is lower than $MQTT\_MAX\_LENGTH\_BYTES$.
Note that, the compiler determined the value of $MQTT\_MAX\_LENGTH\_BYTES$ in the application which is 4 and replaced it (line 27).

\section{CVE Dataset}
\label{sec:appendix-dataset}

% We utilized a CVE dataset comprising 62 real-world CVEs, 
% collected by the state-of-the-art technique, RapidPatch (Table~\ref{table-CVE-dataset-extra}). 
% AutoPatch successfully generated hotpatches for 56 out of the 62 CVEs, 
% demonstrating a success rate exceeding 90\%. 
% CVEs where \SN{} could not generate hotpatches are denoted with '-' in Table~\ref{table-CVE-dataset-extra}, 
% with explanations provided in Section~\ref{sec-eval-effectiveness}.

Table~\ref{tab:CVEdatasetextra} shows the hotpatch execution time and memory overhead 
for the remaining 50 CVEs incurred by \SN{} on the nRF52840 board.
As mentioned in Sections~\ref{sec-eval-effectiveness} and \ref{HED-sec}, \SN{} cannot generate hotpatches for six of the 62 CVEs, 
and six CVEs do not have official patches. 
These are excluded from the evaluation and are marked with "-" in the last three columns.

\begin{table*}[]
    \centering
	\caption{\SN{} results on nRF52840 for the other 50 CVEs. In the "Triggering" column, we use 
	E: function entrance, F: after function call, 
	I: inside/after if statements, and C: inside/after complex loops.
	The first three columns are from RapidPatch~\cite{he2022rapidpatch}.}
    \resizebox{\textwidth}{!}{%
	\begin{footnotesize}
    \begin{tabular}{|c|c|c|c|c|c|}
    \hline
    CVE No.                          & OS/Lib                        & Patch Description / Reason Why Cannot Fix                                                         & Triggering & \begin{tabular}[c]{@{}c@{}}Execution\\ Time ($\mu s$)\end{tabular} & \begin{tabular}[c]{@{}c@{}}Memory\\ Usage (Bytes)\end{tabular} \\ \hline
    2018-16522                       & FreeRTOS                      & Basic block replacement to initialize protocol with memset name buffer after malloc               & I          & 0.65                                                       & 396                                                       \\ \hline
    2018-16526                       & FreeRTOS                      & Rewrite function prvProcessIPPacket to update pxIPHeader-\textgreater{}ucVersionHeaderLength      & F          & 0.18                                                       & 380                                                       \\ \hline
    2018-16525 & FreeRTOS & Add UDP payload length field validation for prvProcessIPPacket function                           & E          & 0.45                                                       & 428                                                       \\ \hline
    2018-16599                       & FreeRTOS                      & Add UDP payload length field validation for prvProcessIPPacket function                           & I          & 0.23                                                       & 432                                                       \\ \hline
    2018-16601                       & FreeRTOS                      & Add IP header length field validation for prvProcessIPPacket function                             & E          & 0.45                                                       & 420                                                       \\ \hline
    2018-16523                       & FreeRTOS                      & Check if uxNewMSS equals zero before use                                                          & I          & 0.31                                                       & 392                                                       \\ \hline
    2018-16602                       & FreeRTOS                      & Add length checks for the DHCP option fields walking loop                                         & I          & 0.37                                                       & 396                                                       \\ \hline
    2018-16600                       & FreeRTOS                      & Check received frame size in function prvProcessEthernetPacket for ARP case                       & I          & 0.31                                                       & 440                                                       \\ \hline
    2018-16527                       & FreeRTOS                      & Check received frame size in function prvProcessIPPacket for ICMP case                            & I          & 0.31                                                       & 440                                                       \\ \hline
    2018-16598                       & FreeRTOS                      & Match outgoing DNS query with received answer                                                     & I          & 0.23                                                       & 380                                                       \\ \hline
    2017-14199                       & ZephyrOS                      & Add out-of-bound check for function dns\_resolve\_cb                                              & I          & 0.35                                                       & 452                                                       \\ \hline
    2017-14201                       & ZephyrOS                      & Modify the way to pass and interpret void *user\_data for function dns\_result\_cb                & E          & 0.48                                                       & 500                                                       \\ \hline
    2017-14202                       & ZephyrOS                      & Macro and struct definitions modification are involved                                            & -          & -                                                          & -                                                         \\ \hline
    2019-9506                        & ZephyrOS                      & Add filter point to ban low encryption key length                                                 & I          & 0.21                                                       & 390                                                       \\ \hline
    2020-10019                       & ZephyrOS                      & Limit upload length to the size of the request buffe                                              & I          & 0.17                                                       & 380                                                       \\ \hline
    2020-10022                       & ZephyrOS                      & Too many vulnerable functions are involved                                                        & -          & -                                                          & -                                                         \\ \hline
    2020-10027                       & ZephyrOS                      & Obtain C flag in APSR after the cmp instruction to help enforce unsigned comparison               & E          & 0.26                                                       & 392                                                       \\ \hline
    2020-10058                       & ZephyrOS                      & Perform validation for arguments of the kscan syscalls                                            & E          & 0.39                                                       & 400                                                       \\ \hline
    2020-10059                       & ZephyrOS                      & Enable DTLS peer checking for the UpdateHub module                                                & E          & 0.23                                                       & 380                                                       \\ \hline
    2020-10060                       & ZephyrOS                      & Add array length check to avoid reference uninitialized stack memory                              & I          & 0.32                                                       & 392                                                       \\ \hline
    2020-10061                       & ZephyrOS                      & Basic block-level prevent referencing uninitialized stack memory in function updatehub\_probe     & I          & 0.23                                                       & 380                                                       \\ \hline
    2020-10064                       & ZephyrOS                      & Patch code is too complex and involves many changes                                               & -          & -                                                          & -                                                         \\ \hline
    2020-10066                       & ZephyrOS                      & Add argument nullpointer checks for function hci\_cmd\_done in Bluetooth HCI core                 & E          & 0.31                                                       & 395                                                       \\ \hline
    2020-10067                       & ZephyrOS                      & Add integer overflow checks for is\_in\_region function                                           & F          & 0.36                                                       & 396                                                       \\ \hline
    2020-10068                       & ZephyrOS                      & Drop reponse with no local initiated request and duplicate requests                               & I          & 0.32                                                       & 392                                                       \\ \hline
    2020-10069                       & ZephyrOS                      & Add arguments validation for function ull\_slave\_setup                                           & F          & 0.39                                                       & 400                                                       \\ \hline
    2020-10070                       & ZephyrOS                      & Add arithmetic overflow and buffer bound checks for function mqtt\_read\_message\_chunk           & E          & 0.23                                                       & 380                                                       \\ \hline
    2020-10071                       & ZephyrOS                      & Further check the length field on publish messages                                                & I          & 0.34                                                       & 396                                                       \\ \hline
    2020-10072                       & ZephyrOS                      & Patch code is too complex and involves many changes                                               & -          & -                                                          & -                                                         \\ \hline
    2020-13598                       & ZephyrOS                      & Macro definition modification are involved                                                        & -          & -                                                          & -                                                         \\ \hline
    2020-13600                       & ZephyrOS                      & Rewrite function eswifi\_reset and \_\_parse\_scan\_res logic to add buffer overflow check        & C          & 0.32                                                       & 392                                                       \\ \hline
    2020-13601                       & ZephyrOS                      & Add out-of-bounds read check in the middle of function dns\_read                                  & I          & 0.45                                                       & 400                                                       \\ \hline
    2020-13602                       & ZephyrOS                      & Basic block-level intercept redundant branch for malformed packet in function do\_write\_op\_tlv  & I          & 0.29                                                       & 388                                                       \\ \hline
    2020-17441                       & PicoTCP                       & No official patch                                                                                 & -          & -                                                          & -                                                         \\ \hline
    2020-17442                       & PicoTCP                       & Validate hop-by-hop IPv6 extension header length field for function pico\_ipv6\_process\_hopbyhop & F          & 0.23                                                       & 380                                                       \\ \hline
    2020-17444                       & PicoTCP                       & No official patch                                                                                 & -          & -                                                          & -                                                         \\ \hline
    2020-24337                       & PicoTCP                       & Validate TCP packet option length field before invoking option handler                            & I          & 0.38                                                       & 398                                                       \\ \hline
    2020-24338                       & PicoTCP                       & Add bound check on iterator for the while loop in function pico\_dns\_decompress\_name            & C          & 0.36                                                       & 396                                                       \\ \hline
    2020-24339                       & PicoTCP                       & Add out-of-bounds check for iterator of pico\_dns\_packet in function pico\_dns\_decompress\_name & I          & 0.36                                                       & 396                                                       \\ \hline
    2020-17437                       & uIP                           & No official patch                                                                                 & -          & -                                                          & -                                                         \\ \hline
    2020-24334                       & uIP                           & No official patch                                                                                 & -          & -                                                          & -                                                         \\ \hline
    2021-3336                        & wolfSSL                       & Exit function DoTls13CertificateVerify on signature without corresponding certificate             & I          & 0.22                                                       & 380                                                       \\ \hline
    2020-24585                       & wolfSSL                       & Add check to reject DTLS application data messages in epoch 0 as out of order                     & I          & 0.75                                                       & 420                                                       \\ \hline
    2020-12457                       & wolfSSL                       & Prevent multiple ChangeCipherSpecs in a row                                                       & I          & 0.23                                                       & 380                                                       \\ \hline
    2019-18840                       & wolfSSL                       & Need to add too many sanity checks                                                                & -          & -                                                          & -                                                         \\ \hline
    2019-16748                       & wolfSSL                       & Add sanity check on length before read for function CheckCertSignature\_ex                        & I          & 0.31                                                       & 392                                                       \\ \hline
    2020-24335                       & Contiki-NG                    & No official patch                                                                                 & -          & -                                                          & -                                                         \\ \hline
    2020-24336                       & Contiki-NG                    & Validate DNS answer‘s length field before using it for memcpy in function ip64\_dns64\_4to6       & I          & 0.35                                                       & 396                                                       \\ \hline
    2020-13987                       & Contiki                       & Add out-of-bounds check for upper\_layer\_len in function upper\_layer\_chksum                    & F          & 0.25                                                       & 390                                                       \\ \hline
    2020-17439                       & Contiki                       & No official patch                                                                                 & -          & -                                                          & -                                                         \\ \hline
    \textbf{Average}                 &                               &                                                                                                   &            & \textbf{0.33}                                              & \textbf{400}                                                          \\ \hline    
\end{tabular}
    \end{footnotesize}
    }
    \label{tab:CVEdatasetextra}
    \end{table*}

\end{document}